\documentclass[aps,preprintnumbers,superscriptaddress,showpacs,nofootinbib,amsmath, amssymb]{revtex4-2}

\usepackage{graphicx}
\usepackage{subfigure}
\usepackage{hyperref}
\usepackage{color}
\usepackage[table]{xcolor}
\usepackage{boldline,multirow}
\usepackage{booktabs}
\usepackage{pifont}
\usepackage{mathrsfs}
\usepackage{soul}
\usepackage[utf8]{inputenc}
\usepackage{multirow}

\usepackage{color}
\definecolor{rosso}{cmyk}{0,1,1,0.4}
\definecolor{rossos}{cmyk}{0,1,1,0.55}
\definecolor{rossoc}{cmyk}{0,1,1,0.2}
\definecolor{blu}{cmyk}{1,1,0,0.3}
\definecolor{blus}{cmyk}{1,1,0,0.6}
\definecolor{bluc}{cmyk}{1,1,0,0.1}
\definecolor{verde}{cmyk}{0.92,0,0.59,0.25}
\definecolor{verdec}{cmyk}{0.92,0,0.59,0.15}
\definecolor{verdes}{cmyk}{0.92,0,0.59,0.4}
\definecolor{bviolet}{rgb}{0.54, 0.17, 0.89}
\hypersetup{colorlinks,bookmarksopen,bookmarksnumbered,citecolor=bviolet,
linkcolor=bviolet,pdfstartview=FitH,urlcolor=bviolet}

\def\apjs{Astrophys.\ J.\ Supp.}
\def\apj{Astrophys.\ J.}

\def\pasp{Pub. Astron. Soc. Pacific}

\newcommand{\rc}{r_\text{c}}

\newcommand{\vmax}{V_{\rm{max}}}
\newcommand{\rmax}{R_{\rm{max}}}

\newcommand{\sigmam}{\sigma/m}

\newcommand{\chisq}{\chi^{2}}

\newcommand{\kms}{\,\mathrm{km/s}}
\newcommand{\cmg}{\,\mathrm{cm^2/g}}

\newcommand{\kpc}{\,\mathrm{kpc}}

\newcommand{\mchi}{m_{\chi}}
\newcommand{\mphi}{m_{\phi}}
\newcommand{\alphachi}{\alpha_{\chi}}
\newcommand{\sigmaeff}{\sigma_{\text{eff}}}
\newcommand{\sigmavel}{\sigma_\text{1D}}
\newcommand{\sigmaeffm}{\sigmaeff/m}

\begin{document}

\title{Gravothermal collapse and the diversity of galactic rotation curves}
\author{M. Grant Roberts}\email{migrober@ucsc.edu}
\affiliation{Department of Physics, 1156 High St., University of California Santa Cruz, Santa Cruz, CA 95064, USA}
\affiliation{Santa Cruz Institute for Particle Physics, 1156 High St., Santa Cruz, CA 95064, USA}
\author{Manoj Kaplinghat}\email{mkapling@uci.edu}
\affiliation{Department of Physics and Astronomy, University of California, Irvine, CA 92697 USA}

\author{Mauro Valli}\email{vallima@roma1.infn.it}
\affiliation{INFN Sezione di Roma, Piazzale Aldo Moro 2, I-00185 Rome, Italy}

\author{Hai-Bo Yu }\email{haiboyu@ucr.edu}
\affiliation{Department of Physics and Astronomy, University of California, Riverside, CA 92521 USA}

\begin{abstract}

The rotation curves of spiral galaxies exhibit a great diversity that challenge our understanding of galaxy formation and the nature of dark matter. Previous studies showed that in self-interacting dark matter (SIDM) models with a cross section per unit mass of $\sigma/m\approx{\cal O}(1)~{\rm cm^2/g}$, the predicted dark matter central densities are a good match to the observed densities in galaxies. In this work, we explore a regime with a larger cross section of $\sigma/m\approx20\textup{--}40~{\rm cm^2/g}$ in dwarf galactic halos. We will show that such strong dark matter self-interactions can further amplify the diversity of halo densities inherited from their assembly history. High concentration halos can enter the gravothermal collapse phase within $10~{\rm Gyr}$, resulting in a high density, while low concentration ones remain in the expansion phase and have a low density. We fit the rotation curves of $14$ representative low surface brightness galaxies and demonstrate how the large range of observed central densities are naturally accommodated in the strong SIDM regime of $\sigma/m\approx20\textup{--}40~{\rm cm^2/g}$. Galaxies that are outliers in the previous studies due to their high halo central densities, are no longer outliers in this SIDM regime as their halos would be in the collapse phase. For galaxies with a low density, the SIDM fits are robust to the variation of the cross section. Our findings open up a new window for testing gravothermal collapse, the unique signature of strong dark matter self-interactions, and exploring a broader SIDM model space. As an example, we illustrate how the larger cross sections favored by our fits, together with upper limits from strong lensing observations in clusters, pick out the preferred SIDM model space for a dark matter particle coupled to a light gauge boson in the Born regime.

\end{abstract}

\maketitle

\section{Introduction}

The rotation curves of spiral galaxies exhibit a great diversity, which is challenging to understand in the standard model of galaxy formation; see~\cite{deBlok:2009sp,McGaugh:2020ppt,Nesti:2023tid} for general reviews on this topic. In particular, Ref.~\cite{KuziodeNaray:2009oon} analyzed the rotation curves of low surface brightness galaxies and found that the inner halo density can vary a factor of $30$ among the galaxies with a similar maximum circular velocity of the halo; see also~\cite{Gentile:2004tb,KuziodeNaray:2006wh,deBlok:2001hbg,deBlok:2001rgg,McGaugh:2001yc}. Ref.~\cite{Oman:2015xda} compiled a large dataset with more than $180$ spiral, dwarf, low and high surface brightness galaxies, and found that the spread in the circular velocity in the inner region ($\sim2~{\rm kpc}$) is a factor of $4$ asymptotic velocities at large radii of 70-80 $\rm km/s$. A similar spread was also noted~\cite{Li:2018tdo,Ren:2018jpt} in the SPARC dataset~\cite{Lelli:2016zqa, SPARCdata}, a sample of $175$ nearby galaxies with surface photometry at $3.6~{\rm \mu m}$ and high-quality rotation curves collected from the literature. More recent measurements of the rotation curves for $26$ dwarf galaxies reveal such a diversity as well~\cite{Relatores:2019ceq,Relatores:2019ews}. The standard cold, collisionless dark matter (CDM) scenario, after properly taking into account baryonic feedback processes associated with galaxy formation, may explain the diversity; see~\cite{Santos_Santos_2017}. Nevertheless, so far, the solution from baryonic physics has not been satisfactory and more work is needed; see~\cite{Santos-Santos:2019vrw,Kaplinghat:2019dhn} for relevant discussions. 

If dark matter has strong self-interactions, its distribution in galactic halos can be more diverse due to collisional thermalization of dark matter particles in the inner regions; see~\cite{Tulin:2017ara,Adhikari:2022sbh} for reviews. The thermalization can lead to cored {\it and} cuspy density profiles, with details depending on the self-interacting cross section, the dark matter halo and the baryonic concentration. Taking these into account, Ref.~\cite{Kamada:2016euw} performed SIDM fits to the rotation curves of $30$ spiral galaxies and Ref.~\cite{Ren:2018jpt} further fitted $135$ galaxies from the SPARC sample and found that SIDM can provide a good fit for most of the galaxies in the sample; see also~\cite{Zentner:2022xux}. These SIDM fits are based on a semi-analytical isothermal halo model~\cite{Kaplinghat:2013xca,Kaplinghat:2015aga}, which has been tested in N-body simulations~\cite{Elbert:2016dbb,Creasey:2017qxc,Sameie:2018chj,Robertson:2020pxj,Jiang:2022aqw}. For galaxies where dark matter dominates the dynamics, the self-interactions produce a shallow density core. As the concentration of baryons increases, the core size decreases, while the core density increases~\cite{Kaplinghat:2013xca}. Additionally, the scatter in the halo concentration--mass relation also plays a vital role in the SIDM fits~\cite{Kamada:2016euw,Creasey:2017qxc}.

The SIDM fits in~\cite{Ren:2018jpt} fix the self-scattering cross section per mass $\sigma/m$ to be $3~{\rm cm^2/g}$. This is approximately the minimal cross section that is needed to fit the data of dwarf galaxies that have the most slowly rising rotation curves in the sample. The fits also assume that the halos are in the expansion phase of the gravothermal evolution. This is justified because almost all halos would be in the expansion phase within $10~{\rm Gyr}$ for the adopted cross section. The fits in~\cite{Ren:2018jpt} are not highly sensitive to the value of cross section and larger values would work just as well. This is because the halo properties do not change rapidly deep in the expansion phase, see, e.g.,~\cite{Elbert:2014bma}. However, with much larger cross sections, the possibility of significant core-collapse in halos~\cite{Balberg:2002ue,Koda:2011yb} opens up and a new analysis is required.

In this work, we motivate the need for large cross sections that allow for core-collapse in addressing the diversity problem of spiral galaxies in SIDM. It is important to note that not all low surface brightness galaxies, where dark matter dominates dynamically in all radii, have low inner dark matter densities. Instead, some of them have surprisingly high densities, as highlighted in~\cite{KuziodeNaray:2009oon,Santos-Santos:2019vrw}. The latest measurements in~\cite{Relatores:2019ceq,Relatores:2019ews} found galaxies that belong to this category. For example, NGC~7320 is extremely dense and its circular velocity reaches the asymptotic value $\sim80~\kms$ at $\sim1.5~{\rm kpc}$. This galaxy is an outlier for SIDM with $\sigma/m\approx{\cal O}(1)~{\rm cm^2/g}$, even after accounting for a possible large spread in the halo concentration--mass
relation. Intriguingly, SIDM has a built-in mechanism to further amplify the diversity. Gravothermal evolution of an SIDM halo can lead to a high inner halo density if the halo enters the collapse phase~\cite{Balberg:2002ue,Koda:2011yb}. This can be achieved within $10~{\rm Gyr}$ for a cross section of $\sigma/m\approx20\textup{--}40~{\rm cm^2/g}$ and higher at velocities comparable to the asymptotic circular velocities in these galaxies. 

Crucially, even in the regime of $\sigma/m\approx20\textup{--}40~{\rm cm^2/g}$ most of the halos would not have undergone significant core-collapse within $10~{\rm Gyr}$. The collapse time is sensitive to the concentration $c_{200}$ as $t_c\propto (\sigma/m)^{-1}c^{-7/2}_{200}M^{-1/3}_{200}$~\cite{Essig:2018pzq,Kaplinghat:2019svz,Sameie:2019zfo,Zeng:2021ldo,Nadler:2023nrd}, and halos with larger concentrations would be further into the collapse phase and hence have a denser core. Halos with lower concentration would not have had time to get deeply into the collapse phase and hence can have a low density, preserving the successes of the SIDM fits found for models with $\sigma/m\approx{\cal O}(1)~{\rm cm^2/g}$ as in~\cite{Ren:2018jpt}. Thus, the diversity encoded through the scatter in the halo concentration can be further amplified for large cross sections that can leads to core-collapse, as demonstrated in recent cosmological N-body simulations, see, e.g.~\cite{Correa:2022dey,Yang:2022mxl,Nadler:2023nrd}.

To demonstrate these points concretely, we choose $14$ representative low surface brightness galaxies, which have high-quality rotation curves, and perform SIDM fits. These galaxies were chosen to be dark matter-dominated to minimize the dynamical impact of the baryons on the SIDM density profile. We apply a parametric SIDM halo model~\cite{Yang:2023jwn,Yang:2024uqb} to fit the rotation curves. The model takes into account the full gravothermal history of a halo and it provides a unified way to model the density profile of an SIDM halo at any given phase of its evolution. Additionally, we cross-check the density profile of a collapsed SIDM halo using an analytical form motivated by conducting fluid simulations in Ref.~\cite{Jiang2021}. The rest of the paper is organized as follows. In Sec.~\ref{sec:sample}, we discuss the galaxies that we consider in this work, and provide details about the parametric SIDM halo model, the numerical routine, and discuss a separate analytical core-collapse density profile. In Sec.~\ref{sec:results}, we discuss the main results inferred from the fits, including various correlation relations among the halo parameters. We conclude in Sec.~\ref{sec:con}. In Appendix~\ref{sec:app-ccp-density}, we present the details of the analytical density profile for halos in the core-collapse phase. In Appendices~\ref{sec:app-contour-plots}, \ref{sec:app-parameters-table}, and~\ref{sec:app-rotation-curves}, we provide additional details about the fits.

\section{Example rotation curves and the parametric SIDM model}
\label{sec:sample}

\subsection{Galaxy sample}
\label{sec:galaxy-sample}

We employ general guidelines for the selection of our sample galaxies from the SPARC dataset~\cite{Lelli:2016zqa, SPARCdata} and Refs.~\cite{Relatores:2019ceq,Relatores:2019ews}. We perform an initial fit to the rotation curves of all of the galaxies within SPARC, and we vary the stellar mass to light ratios of the disk and the bulge, $\Upsilon_{\star,\rm{disk}}$ and $\Upsilon_{\star,\rm{bulge}}$, respectively, using a uniform prior between $0.2~M_{\odot}/L_{\odot}$ and $0.8~M_{\odot}/L_{\odot}$. We remove the galaxies that have a significant dynamical contribution from stars and gas in the flat part of the rotation curve (as typically above $50\%$ of the total rotation curve) and $\vmax \gtrsim 150 \kms$. We then select the the most baryon poor objects out of the remaining
set of galaxies. In particular, we choose galaxies that are dark matter-dominated at their innermost measured data point, typically about $0.5\textup{--}1 \kpc$. The galaxies thus chosen are: DDO~154, ESO~444-G084, F568-V1, F574-1, NGC~0300, NGC~3109, UGC~00128, UGC~06446, UGC~06667, and UGC~08286. 

Refs.~\cite{Relatores:2019ceq,Relatores:2019ews} presented high resolution H$\alpha$ kinematics for $26$ dwarf galaxies. They split the sample into three categories (grade-1, grade-2, and grade-3) based on the quality of the model fit to the data, as well as the strength of radial motion of the gas. The sample is made of $18$ grade-1 and grade-2 galaxies, and $8$ grade-3 galaxies. Of the $18$ acceptable galaxies, we select four from the grade-1 category, which are the most baryon poor. The galaxies thus chosen are: NGC~4376, NGC~4396, NGC~7320, and UGC~4169. Each galaxy is dark matter-dominated at $r \leq 0.5 \kpc$.

\begin{figure}[h]
\includegraphics[scale=0.8]{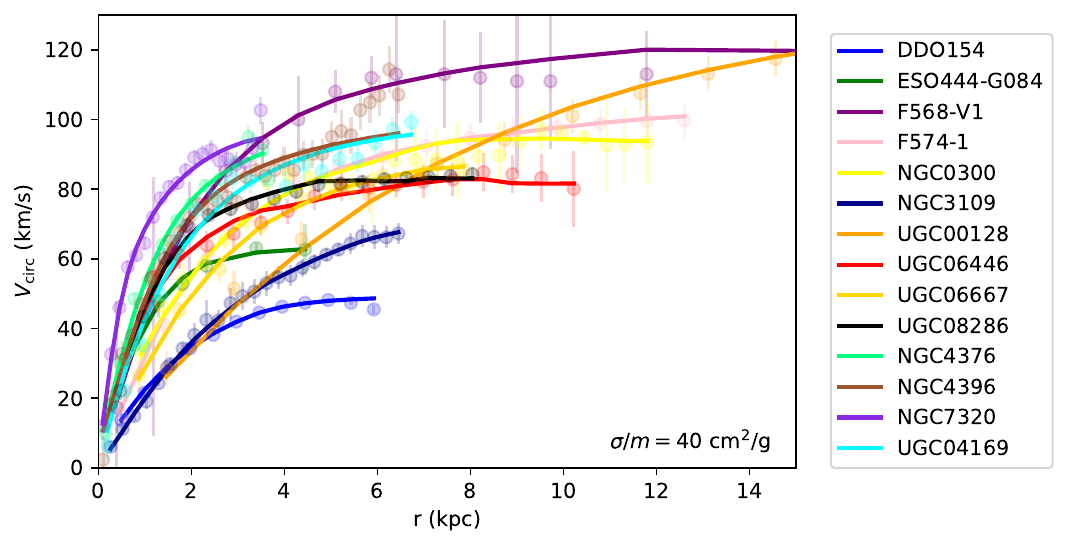}
\caption{The circular velocity profiles of spiral galaxies analyzed in this work (dots with error bars), as well as their corresponding best SIDM fits assuming a self-interacting cross section per particle mass of $\sigma/m=40~{\rm cm^2/g}$ (solid curves).}
\label{fig:combo}
\end{figure}

Fig.~\ref{fig:combo} shows the rotation curves of the spiral galaxies that we consider in this work. Within the sample, the asymptotic rotation speed spans from $40~{\rm km/s}$ to $120~{\rm km/s}$, and most of them are in the $60\textup{--}80~{\rm km/s}$ range. For this sample, the spread in the rotation speed at $2~{\rm kpc}$ is a factor of $4$. Since the galaxies in the sample are dark matter-dominated in all radii, the large spread in $V_{\rm circ}$ in the inner regions indicates that dark matter densities there are diverse. We will show such a diversity can be better explained in SIDM with a cross section of $\sigma/m\approx20\textup{--}40~{\rm cm^2/g}$ than the assumption $\sigma/m\approx{\cal O}(1)~{\rm cm^2/g}$.

\subsection{A parameteric SIDM halo model}
\label{subsec:parametric-model}

We use the parametric SIDM halo model, which was proposed and tested in~\cite{Yang:2023jwn,Yang:2024uqb}, to fit the rotation curves. The model uses a functional form to describe the density profile of an SIDM halo over its entire evolution history. In this work, we will take the density profile (Eq. A.1 of~\cite{Yang:2023jwn}) based on the ``Read" profile~\cite{Read:2015sta}, which is formulated using the Navarro-Frenk-White (NFW) density profile~\cite{Navarro:1996gj},
\begin{eqnarray}
\label{eq:read}
\rho_{\rm Read}(r) = f(r) \rho_{\rm NFW}(r) + \frac{1-f^2(r)}{4\pi r^2 r_c } M_{\rm NFW}(r),
\end{eqnarray}
    where $r_c$ is the core radius, and $f(r) \equiv \tanh(r/r_c)$. The NFW density and mass profiles are 
\begin{eqnarray}
\label{eq:nfw}
\rho_{\rm NFW}(r) = \frac{\rho_s r^3_s}{r \left(r +r_s\right)^2},~
M_{\rm NFW}(r) = 4\pi \rho_s r_s^3 \left[ \ln\left(1+\frac{r}{r_s}\right) -\frac{r}{r+r_s} \right],
\end{eqnarray}
respectively. The enclosed mass follows the relation $M_{\rm Read} = f(r) M_{\rm NFW}(r)$. For $r\gg r_c$, $f(r)=1$, and $M_{\rm Read} = M_{\rm NFW}(r)$. In the limit of $r_c\rightarrow0$, $f\rightarrow1$ and $\rho_{\rm Read}(r)\rightarrow\rho(r)_{\rm NFW}$.

Ref.~\cite{Yang:2023jwn} obtained the evolution of the model parameters in Eqs.~\ref{eq:read} and \ref{eq:nfw}: $\rho_s$, $r_s$, and $r_c$ by fitting to a simulated SIDM halo in~\cite{Yang:2022hkm}
\begin{eqnarray}
\label{eq:m0read}
\frac{\rho_s}{\rho_{s,0}} &=& 1.335 + 0.7746 \tau + 8.042 \tau^5 -13.89 \tau^7  + 10.18 \tau^9 + (1-1.335) (\ln 0.001)^{-1} \ln \left( \tau + 0.001 \right), \nonumber \\
\frac{r_s}{r_{s,0}} &=& 0.8771 - 0.2372 \tau +  0.2216 \tau^2 -0.3868 \tau^3 + (1-0.8771) (\ln 0.001)^{-1} \ln \left( \tau + 0.001 \right), \nonumber \\
\frac{r_c}{r_{s,0}} &=& 3.324 \sqrt{\tau} -4.897 \tau + 3.367 \tau^2 -2.512 \tau^3 + 0.8699 \tau^4,  
\end{eqnarray}
where the subscript $``0"$ denotes the corresponding value of the initial NFW profile, and $\tau=t/t_c$ is the normalized evolution time of the halo with respective to the collapse time~\cite{Balberg:2002ue} 
\begin{equation}
\label{eq:tc0}
t_c=\frac{150}{C} \frac{1}{(\sigma/m) \rho_{s,0} r_{s,0}} \frac{1}{\sqrt{4\pi G \rho_{s,0}}},
\end{equation}
where $\sigma/m$ is the self-scattering cross section per particle mass, $C$ is a constant that can be calibrated with N-body simulations~\cite{Koda:2011yb,Pollack:2014rja,Essig:2018pzq,Nishikawa:2019lsc,Yang:2022zkd}, and we fix it to $C=0.75$. In this work, we set $t=t_{\rm age}=10~{\rm Gyr}$, an approximation for the age of the spiral galaxies in our sample $t_{\rm age}$. Although $t_{\rm age}$ may vary from $10~{\rm Gyr}$ for individual galaxies, a small variation does not affect our SIDM fits and inferences due to degeneracy among the model parameters $\sigma/m$, $t_{\rm age}$, and $c_{200}$ \cite{Yang:2023jwn}. Furthermore, $\sigma/m$ should be regarded as the effective cross section for a given halo mass~\cite{Outmezguine:2022bhq,Yang:2022hkm,Yang:2022zkd}, i.e., the velocity-averaged total cross section weighted by a factor $v^5\sin^2\theta$, where $v$ is the relative velocity of dark matter particles and $\theta$ is the scattering angle. The weighting factor is motivated by the calculation of heat conductivity in kinetic theory. In general, $\sigma/m$ is velocity-dependent and varies with the halo mass. However, it is challenging to infer the velocity dependence of the cross section for individual galaxies due to the degeneracy in the fits; see~\cite{Ren:2018jpt} for detailed discussion. For simplicity, we take four constant values for the cross section for the whole sample $\sigma/m=0~{\rm cm^2/g}$, i.e., the NFW halo limit, $3~{\rm cm^2/g}$, $20~{\rm cm^2/g}$, and $40~{\rm cm^2/g}$.   

In addition to the parametric model presented above, which was calibrated to N-body simulations, we also cross-check the SIDM fits with an analytical core-collapse profile which is constructed to meet the typical core-collapse criteria found in conducting fluid simulations~\cite{Balberg:2002ue,Outmezguine:2022bhq}: ${d\rho(r)}/{dr} = 0\ \text{as}\  r\rightarrow 0$, ${d^2\rho(r)}/{dr^2}\propto -{\rho(0)}/{\rc^2}\ \text{as}\ r\rightarrow 0$, and ${d\ln(\rho)}/{d\ln(r)} = -\alpha\ \text{outside the core}\ \text{with}\ \alpha=2.19$. These conditions dictate the structure of the central core of a halo in the core-collapse phase and the profile outside the core, and they guide the functional form we expect for the density profile in the collapse phase. In this work we consider an analytical core-collapse profile as $\rho(x) \propto \tanh(x/x_{c})^\alpha/(x^\alpha(b+x)^{3-\alpha})$ for a halo in the core-collapse phase, where $x\equiv r/r_{s,0}$, $x_{c}\equiv r_{c}/r_{s,0}$, and $b$ is a numerical factor, typically around $3\textup{--}4$, see Appendix~\ref{sec:app-ccp-density} for full details. Note that in the limit of large radii, the analytical profile above goes like $1/x^{3}$, which is similar to the Read profile, as both map onto the NFW profile in this region. 

Since $\alpha \neq 1$ for the analytical core-collapse profile, it has a mildly different logarithmic slope compared to the Read profile in the intermediate region between the core and the outer NFW-like halo. Despite the difference, we have checked that the fits using the parametric model and the analytical core-collapse profile are almost identically for $\sigma/m=20~{\rm cm^2/g}$ and $40~{\rm cm^2/g}$. Note we do not apply the analytical profile to $\sigma/m=3~\cmg$, as it is purposely constructed for core-collapse halos. In the rest of the paper, we will mainly show the results based on the parametric model. In Sec.~\ref{sec:log-slopes}, we will discuss the logarithmic density slope of NGC~7320 from the fit using the analytical core-collapse profile and show explicitly the similarity in the fits of all galaxies in the sample.

\subsection{The MCMC analysis}

We calculate the circular velocity as
\begin{equation}
V^2_{\rm circ}(r) = V^2_{\rm halo}(r)+V^2_{\rm gas}(r)+\Upsilon_{\star,\rm{disk}}V^2_{\rm disk}(r),
\end{equation}
where $V_{\rm halo}$, $V_{\rm gas}$, $V_{\rm disk}$ are halo, gas, and disk contributions, respectively. In our sample, all galaxies are bulgeless. In our numerical analysis, we consider two sets of fixed stellar mass-to-light ratios $\Upsilon_{\star,\rm{disk}} = 0.5~ M_{\odot}/L_{\odot}$ and $1~ M_{\odot}/L_{\odot}$, as well as varying ones within a range of $0.2\textup{--}0.8~M_{\odot}/L_{\odot}$ with a flat prior. Our SIDM fits are robust to these choices and the results do not change quantitatively. This is not surprising, as they are dark matter-dominated. For the main results, we will report the fits with $\Upsilon_{\star,\rm{disk}}  = 0.5~ M_{\odot}/L_{\odot}$ motivated as follows. For the SPARC dataset, the mean value is $\Upsilon_{\star,\rm{disk}} \approx 0.5~ M_{\odot}/L_{\odot}$. Additionally, Ref.~\cite{Ren:2018jpt} found that the distribution of $\Upsilon_{\star,\rm{disk}}$ is peaked at $\Upsilon_{\star,\rm{disk}} = 0.5~ M_{\odot}/L_{\odot}$ and falls sharply past $0.8~ M_{\odot}/L_{\odot}$ even the assuming a wide prior $0.1\textup{--}10~M_{\odot}/L_{\odot}$. In Appendix~\ref{sec:app-contour-plots}, we include the results with varying $\Upsilon_{\star,\rm{disk}} = 0.2 \textup{--} 0.8~ M_{\odot}/L_{\odot}$ for comparison.

For the halo component, via a log-uniform prior distribution the scale density of the initial NFW halo in a range of $\rho_{s,0}=3.65\times10^5\textup{--}2\times10^9~M_{\odot}/\rm{kpc^{3}}$ and the scale radius $r_{s,0}=1\textup{--}100~{\rm kpc}$. We also impose a top-hat prior on the concentration--mass relation $c_{200}\textup{--}M_{200}$ from~\cite{Diemer2019}, with a width of $0.55$ dex which corresponds to $\pm5\sigma$ scatter. We use the well-known sampler~\texttt{emcee}~\cite{emcee} and the package~\texttt{Colossus}~\cite{colossus}.

\section{Results}
\label{sec:results}

\subsection{Two representative examples}

\begin{figure*}[h]
\includegraphics[scale=1.2]{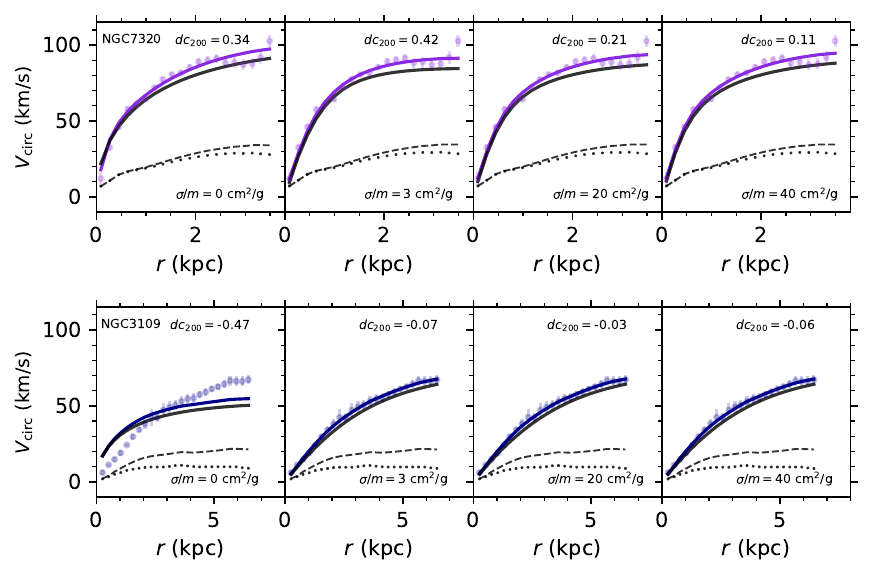}
\caption{The best model fits (solid) to the observed rotation curve (dots with error bars) of NGC 7320 (top) and NGC 3109 (bottom) for $\sigma/m=0~{\rm cm^2/g}$, $3~{\rm cm^2/g}$, $20~{\rm cm^2/g}$, and $40~{\rm cm^2/g}$ from left to right panels, including the halo contribution (solid-black) and the total baryon contribution (dashed-black); the stellar contribution is indicated as well (dotted-black). In each panel, the halo concentration, normalized to the median $dc_{200}=\log_{10}(c_{200}/c^{\rm med}_{200})$, is shown, and the $1\sigma$ scatter is $0.11$ dex.} 
\label{fig:example}
\end{figure*}

We first take NGC~7320 and NGC~3109 to highlight the SIDM solution to the diversity problem. Fig.~\ref{fig:example} (top) shows the observed rotation curve of NGC~7320 (colored dots with error bars) and the model fits (solid-colored) for $\sigma/m=0~{\rm cm^2/g}$, $3~{\rm cm^2/g}$, $20~{\rm cm^2/g}$, and $40~{\rm cm^2/g}$ from left to right panels, including the halo contribution (solid-black) and the total baryon contribution (dashed-black); the stellar contribution is indicated as well (dotted-black). The rotation curve of NGC~7320 in the central regions is well resolved and it rises extremely fast and the circular velocity reaches $V_{\rm circ}\approx60~{\rm km/s}$ at $r=1~{\rm kpc}$. This indicates the halo is extremely dense, as the baryon contribution is negligible. In the limit $\sigma/m=0~{\rm cm^2/g}$, an NFW halo with $V_{\rm max}\approx80~{\rm km/s}$ can provide an excellent fit, but the required concentration compared to the median concentration for a given halo mass, $c^{\rm med}_{200}$, is high. We define the quantity, $dc_{200}\equiv\log_{10}(c_{200}/c^{\rm med}_{200})$, to normalize each halo with respect to the corresponding median concentration to better compare the spread in concentration across our sample. With the NFW fit, NGC~7320 has $dc_{200}=0.34$, which is $3\sigma$ higher than the cosmological median. When $\sigma/m$ increases to $3~{\rm cm^2/g}$, the fit is still excellent $(\min \chisq/\text{d.o.f.} \approx 1.28)$, but the concentration needs to be even higher $dc_{200}=0.42$. For $\sigma/m=3~{\rm cm^2/g}$, we cannot even find a fit if the halo is in the expansion phase, and a high concentration of $dc_{200}=0.42$ is needed such that it can be close to the collapse phase within $t_{\rm age}\approx10~{\rm Gyr}$. As the cross section further increases to $\sigma/m=20~{\rm cm^2/g}$ and $40~{\rm cm^2/g}$, the required concentration decreases to $dc_{200}=0.21$ and $0.11$, respectively, as expected from the scaling relation $t_c\propto(\sigma/m)^{-1}c^{-7/2}_{200}M^{-1/3}_{200}$. Thus NGC~7320 might be an outlier with respective to the concentration-mass relation for $\sigma/m\lesssim3~{\rm cm^2/g}$, but it would be not be an outlier for $\sigma/m\approx20\textup{--}40~{\rm cm^2/g}$.

Fig.~\ref{fig:example} (bottom) shows the rotation curve of NGC~3109 and the model fits. In contrast to NGC~7320, NGC~3109 is in the opposite limit as its rotation curve rises very slowly. At $r\approx1~{\rm kpc}$, $V_{\rm circ}\approx10~{\rm km/s}$, a factor $6$ smaller than that of NGC~7320. In fact, $V_{\rm circ}(r)\propto r$ within $5~{\rm  kpc}$, indicating that the halo has a $5~{\rm kpc}$ core. It is not surprising that an NFW profile cannot give rise to a fit, but all three SIDM fits are excellent (see Table~\ref{tab:numerics2}). In contrast to the case of NGC~7320, the halo concentration is close to the median and it varies very mildly with the cross section; we will come back this point later. NGC~3109 has $V_{\rm max}\approx70~{\rm km/s}$ and NGC~7320 has $V_{\rm max}\approx80~{\rm km/s}$, and hence their halo masses are similar. SIDM with $\sigma/m=3~{\rm cm^2/g}$ can fit both galaxies well, but the spread in the concentration is quite large; NGC~3109 and NGC~7320 have an approximately $0.5$ dex difference in $dc_{200}$. When the cross section increases to $\sigma/m=20~{\rm cm^2/g}$ and $40~{\rm cm^2/g}$, the spread shrinks and the fits are much more in line with each other, the difference in $dc_{200}$ is now only about $0.24$ dex and $0.17$ dex, respectively. Thus strong dark matter self-interactions further amplify the diversity in the density which is inherited from assembly history of the halo. 

If we further increase $\sigma/m>40~{\rm cm^2/g}$, NGC~7320 would require a median halo. In this regard, $\sigmam\approx40~{\rm cm^2/g}$ seems close to an upper limit on the cross section for the relevant mass scale $V_{\rm max}\sim80~{\rm km/s}$, beyond which the majority of halos in the field would be collapsed within the Hubble time, potentially in contradiction to observations of shallowly rising rotation curves. As shown in Fig.~\ref{fig:combo}, many of the galaxies in our sample have a slower rising rotation curve compared to NGC 7320. A dedicated analysis would be required to firm up this upper limit, and we would also need to further validate the analytic model with more simulations to make this upper limit robust. Furthermore, we need a much larger sample of spiral galaxies that are dark matter-dominated, and we will leave it for future work.

\subsection{Halo concentration}

\begin{figure*}[h]
\includegraphics[scale=0.8]{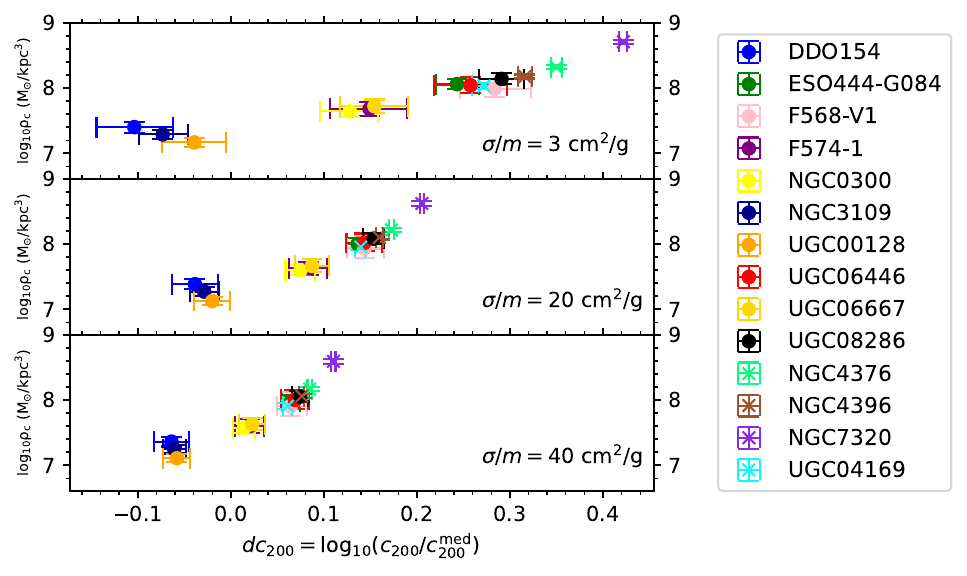}
\caption{The core density vs concentration normalized to the cosmological median from the SIDM fits $\sigma/m=3~{\rm cm^2/g}$ (top), $20~{\rm cm^2/g}$ (middle), and $40~{\rm cm^2/g}$ (bottom). The $1\sigma$ scatter of the concentration is $0.11$ dex.}
\label{fig:rhocdc200}
\end{figure*}

As we demonstrated with NGC~7320 and NGC~3109, strong dark matter self-interactions can diversify the halo density and explain the opposite extremes. In this section, we present the halo core density and concentration for all galaxies in our sample. Since the NFW profile cannot provide a good fit for many of them, we do not show the fits with $\sigma/m=0~{\rm cm^2/g}$.       

Fig.~\ref{fig:rhocdc200} shows the halo core density vs concentration for all galaxies in our sample from the SIDM fits (colored symbols) with $\sigma/m=3~{\rm cm^2/g}$ (top), $20~{\rm cm^2/g}$ (middle), and $40~{\rm cm^2/g}$ (bottom). The concentration is normalized with respect to the cosmological median for a given halo mass, i.e., $dc_{200}=\log_{10}(c_{200}/c^{\rm med}_{200})$. We see that overall the spread in the concentration shrinks as the cross section increases, i.e., $dc_{200}\approx[-0.1,0.44]$,  $[-0.04,0.2]$, and $[-0.06,0.1]$, for $\sigma/m=3~{\rm cm^2/g}$, $20~{\rm cm^2/g}$, and $40~{\rm cm^2/g}$, respectively. For $\sigma/m=3~{\rm cm^2/g}$, eight galaxies would be $2\sigma$ outliers in concentration. When the cross section increases to $\sigma/m=20\textup{--}40~{\rm cm^2/g}$, all the galaxies in the sample are within the $2\sigma$ range around the median. For $11$ galaxies in the sample, their halo concentration decreases as the cross section increases, following a trend as of NGC~7320. On the hand, for DDO~154, NGC~3109, and UGC~00128, the change in the concentration is very mild as we vary the cross section, but notable. Interestingly, their concentration increase slightly, and then decreases. This trend is well expected from gravothermal evolution of an SIDM halo, as we discuss next.

\subsection{Gravothermal evolution}

\begin{figure}[h]
\includegraphics[scale=0.8]{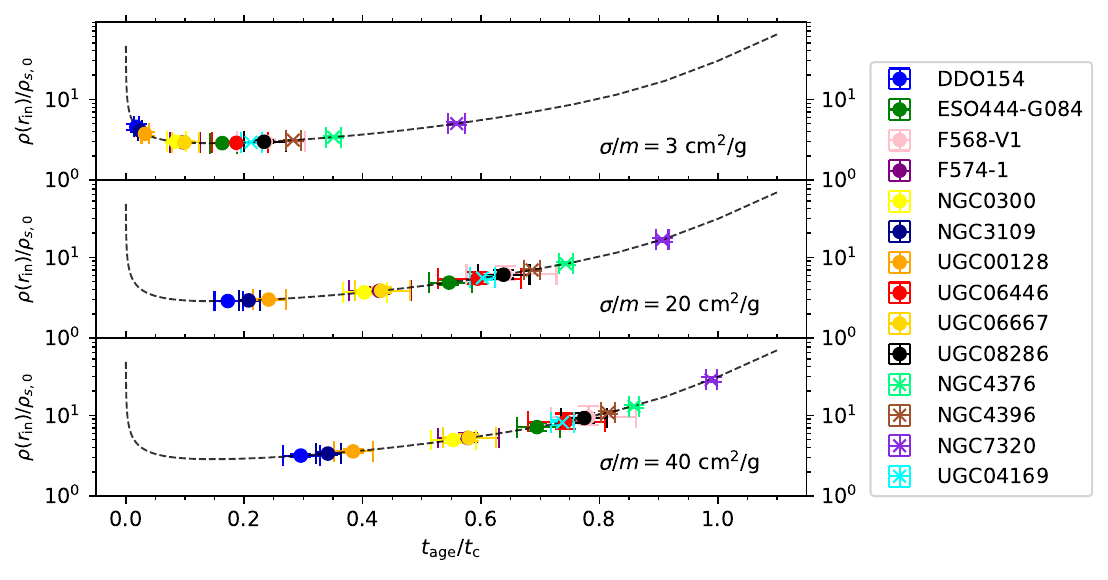}
\caption{The central density vs gravothermal time from the SIDM fits $\sigma/m=3~{\rm cm^2/g}$ (top), $20~{\rm cm^2/g}$ (middle), and $40~{\rm cm^2/g}$ (bottom). The central density, evaluated at the radius $r_{\text{in}} = 10^{-3}r_{s,0}$, is normalized to the scale radius of its corresponding initial NFW halo, and $t_{\rm age}=10~{\rm Gyr}$, and $t_c$ is calculated using Eq.~\ref{eq:tc0}. The dashed curve denotes the universal evolution trajectory of an SIDM halo characterized by Eq.~\ref{eq:m0read}. }
\label{fig:rhotau}
\end{figure}

In Fig.~\ref{fig:rhotau}, we show the central density vs the evolution time for each galaxy from the SIDM fits (colored symbols) $\sigma/m=3~{\rm cm^2/g}$ (top), $20~{\rm cm^2/g}$ (middle), and $40~{\rm cm^2/g}$ (bottom). The central density is evaluated at $r_{\rm{in}}=10^{-3}~r_{s,0}$, and it is further normalized with respect to the scale radius of the initial NFW halo $\rho_{s,0}$. The evolution time is normalized using the collapse timescale $t_c$ in Eq.~\ref{eq:tc0}, i.e., $t_{\rm age}/t_c$ and we have set $t_{\rm age}=10~{\rm Gyr}$ for all galaxies in the sample. Therefore $t_{\rm age}/t_c$ characterizes the stage of gravothermal evolution of the halo for a galaxy observed at present. For comparison, we also show the evolution of the normalized central density in $\tau$ (dashed-black); recall $\tau=t/t_c = t_{\rm age}/t_c$. It is important to note that the $\rho(r_{\rm in})/\rho_{s,0}\textup{--}\tau$ relation is universal and completely determined by Eq.~\ref{eq:m0read}. From the relation, we can show that at $\tau\approx0.146$ an SIDM halo reaches the phase of maximum expansion and its central density is lowest $\rho(r_{\rm in})/\rho_{s,0}\approx2.89$. Afterwards, the central density increases, but very slowly, until $\tau>0.8$. Thus, technically speaking, the collapse starts at $\tau\approx0.146$, the central density only increases substantially when $t_c$ is close to $10~{\rm Gyr}$; though, if any galaxy undergoes a merger, this could reset its position along the curve.   

As the cross section increases, all the galaxies shift towards the collapse phase and $t_{\rm age}/t_c$ increases. For $\sigma/m=3~{\rm cm^2/g}$, the halo of NGC 7320 is mildly collapsed to explain its sharp rising circular velocity as $t_{\rm age}/t_c\approx0.55$. It further increases to $\approx0.9$ and $1$ for $\sigma/m=20~{\rm cm^2/g}$ and $40~{\rm cm^2/g}$, respectively. Since $t_{\rm age}=10~{\rm Gyr}$ is fixed, $t_c$ decreases as $\sigma/m$ increases, and thus the galaxies will move rightward along the relation. On the other hand, NGC~3109 is still in the expansion phase for $\sigma/m=3~{\rm cm^2/g}$ as $t_{\rm age}/t_c\approx0.025$, at which its central density has not reached the minimum yet. For $\sigma/m=20~{\rm cm^2/g}$, $t_{\rm age}/t_c\approx0.2$, the concentration increases mildly as indicated in Fig.~\ref{fig:rhocdc200}. This is because in the expansion phase, a larger cross section tends to create a larger core with lower density, thus a halo with a relatively higher concentration is need to fit the data. For $\sigma/m=40~{\rm cm^2/g}$, NGC~3109 is in the phase of mild collapse as $t_{\rm age}/t_c\approx0.35$. In this case, a larger cross section leads to a smaller core and higher density, hence the concentration needs to shift lower to compensate. We see that DDO~150 and UGC~00128 follow the same trend as NGC~3109, as they have the lowest densities in the sample; see Fig.~\ref{fig:rhocdc200}. We also emphasize that despite the cross section being up to $40~\cmg$, with the exception of NGC 7320, the galaxies in our sample still mostly lie within the middle of the core-collapse phase, rather than deep into the collapse. In general, we do not expect most galaxies to have entered into the deep core-collapse phase within the Hubble time.

Using a parametric model, we have fitted the rotation curves in a manner that continuously traces the gravothermal evolution of SIDM halos. This approach is crucial. For outlier galaxies with sharply rising rotation curves, a pure core-expansion solution may not yield satisfactory fits, even when allowing for large halo concentrations. For instance, in the case of NGC7320, we have verified that an SIDM model with $\sigma/m = 3~{\rm cm^2/g}$ fails to provide a good fit in the core-expansion limit, even when treating $c_{200}$ as a free parameter; see~\cite{Kong:2025irr} for a discussion of similar galaxies recently identified. Moreover, we find that for a small subset of galaxies analyzed in~\cite{Ren:2018jpt}, SIDM fits using the core-expansion solution with $\sigma/m = 3~{\rm cm^2/g}$ tend to underestimate the circular velocities in the central regions, despite allowing the concentration to rise up to $3\sigma$ above the median of the expected distribution. It would be valuable to reanalyze these outliers using the evolutionary approach developed in this work. On the other hand, for galaxies in~\cite{Ren:2018jpt} that favor large central density cores, our current method will yield fits that are nearly identical to those obtained from the core-expansion approach.

Additionally, when the cross section increases to $\sigma/m \sim 20~{\rm cm^2/g}$, the required concentration becomes more consistent with that predicted by the concentration–mass relation from standard cosmological structure formation theory. This alleviates—or even resolves—the “fine-tuning” problem present in both the CDM model and the SIDM model that rely solely on the core-expansion solution. A larger galaxy sample will be essential to further determine the cross section required to bring the outliers into agreement with expectations from cosmic structure formation. 

\subsection{Correlations}
\begin{figure}[h]
\includegraphics[scale=0.8]{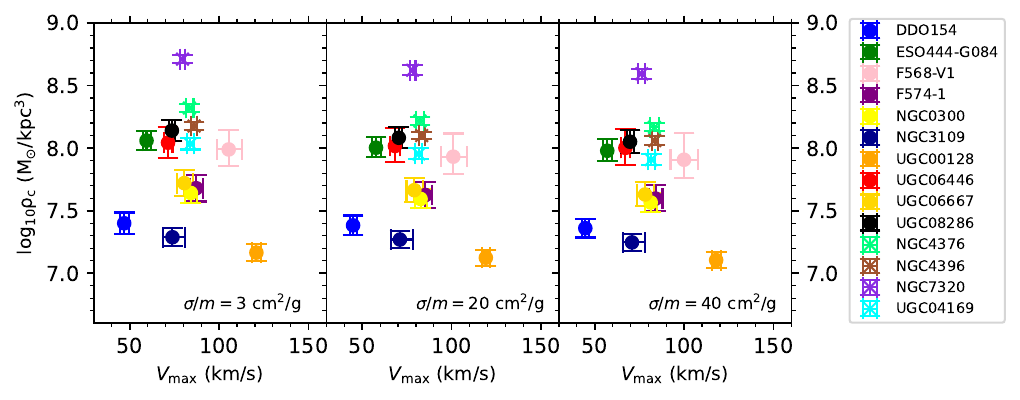}
\caption{The core density vs maximum halo circular velocity from the SIDM fits with $\sigma/m=3~{\rm cm^2/g}$ (left), $20~{\rm cm^2/g}$ (middle), and $40~{\rm cm^2/g}$ (right).}
\label{fig:rhocvmax}
\end{figure}

Fig.~\ref{fig:rhocvmax} shows the core density vs maximum circular velocity $\rho_c\textup{--}V_{\rm max}$ for the galaxies inferred from the SIDM fits $\sigma/m=3~{\rm cm^2/g}$ (left), $20~{\rm cm^2/g}$ (middle), and $40~{\rm cm^2/g}$ (right). Overall, the $\rho_c\textup{--}V_{\rm max}$ trend is not sensitive to the cross section, as expected.  We clearly see the diversity of the core density for given $V_{\rm max}$. For the sample we consider, the spread in $\rho_c$ is largest for $V_{\rm max}\approx70\textup{--}80~{\rm km/s}$, a factor of $25$, with NGC 7320 being densest and NGC 3109 shallowest. Ref.~\cite{KuziodeNaray:2009oon} found a similar spread in the $\rho_c\textup{--}V_{\rm max}$ trend when analyzing a sample of $9$ low surface brightness galaxies and showed that it is puzzling to explain in dark matter models that predict a shallow density core. In our study, although $\sigma/m=3~{\rm cm^2/g}$ can give rise to a fit, the spread in the concentration is large (about a $0.5$ dex difference in $dc_{200}$). When the cross section is increased to $\sigma/m=20\textup{--}40~{\rm cm^2/g}$, dense halos, such as NGC~7320, can be deep in the collapse phase with a relatively smaller concentration which is closer to the median, and the SIDM explanation becomes more natural, because the overall tension in the outliers from the median concentration is reduced significantly with higher cross sections. 
  
\begin{figure}[h]
\includegraphics[scale=0.8]{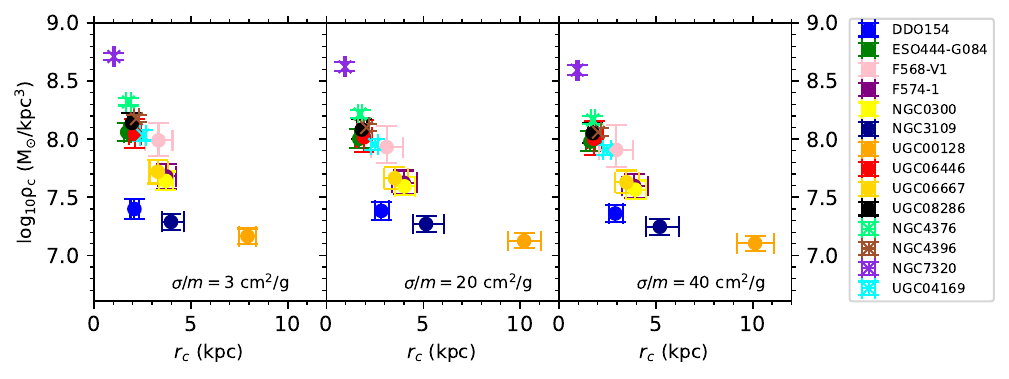}
\caption{The core density vs core radius from the SIDM fits with $\sigma/m=3~{\rm cm^2/g}$ (left), $20~{\rm cm^2/g}$ (middle), and $40~{\rm cm^2/g}$ (right).}
\label{fig:rhocrc}
\end{figure}

Fig.~\ref{fig:rhocrc} shows the core density vs the core radius $\rho_c\textup{--}r_c$ inferred from the SIDM fits with $\sigma/=3~{\rm cm^2/g}$ (left), $20~{\rm cm^2/g}$ (middle), and $40~{\rm cm^2/g}$ (right). Similar to the $\rho_c\textup{--}V_{\rm max}$ trend, the $\rho_c\textup{--}r_c$ trend is insensitive to the cross section. Clearly, there is an anti-correlation between $\rho_c$ and $r_c$. Among the galaxies in the sample, NGC~7320 has the highest density and smallest core size $r_c\approx1~{\rm kpc}$. In contrast, the core size of NGC~3109 is close to $5~{\rm kpc}$. UGC~00128 has the most massive halo $V_{\rm max}\approx120~{\rm km/s}$ and the largest core in the sample, $r_c\approx10~{\rm kpc}$. On the other hand, as indicated in Fig.~\ref{fig:rhocvmax}, F568-V1's $V_{\rm max}\approx100~{\rm km/s}$ is only $20\%$ smaller than UGC~00128's, but the former has a core size of $r_c\approx3~{\rm kpc}$, a factor of $3$ smaller than the latter, while the core density is factor of $10$ higher. Thus the galaxies exhibit the diversity in both core density {\it and} core size. After taking to account full gravothermal evolution of the halo via the parametric model, all three SIDM cross sections can give rise to excellent fits to the data, while $\sigma/m\approx20\textup{--}40~{\rm cm^2/g}$ provides a minimal explanation to the diversity, as the inferred halo concentration is well within the range expected from cosmic structure formation.

\subsection{Logarithmic slopes of density profiles}
\label{sec:log-slopes}

\begin{figure*}[h]
\includegraphics[scale=0.8]{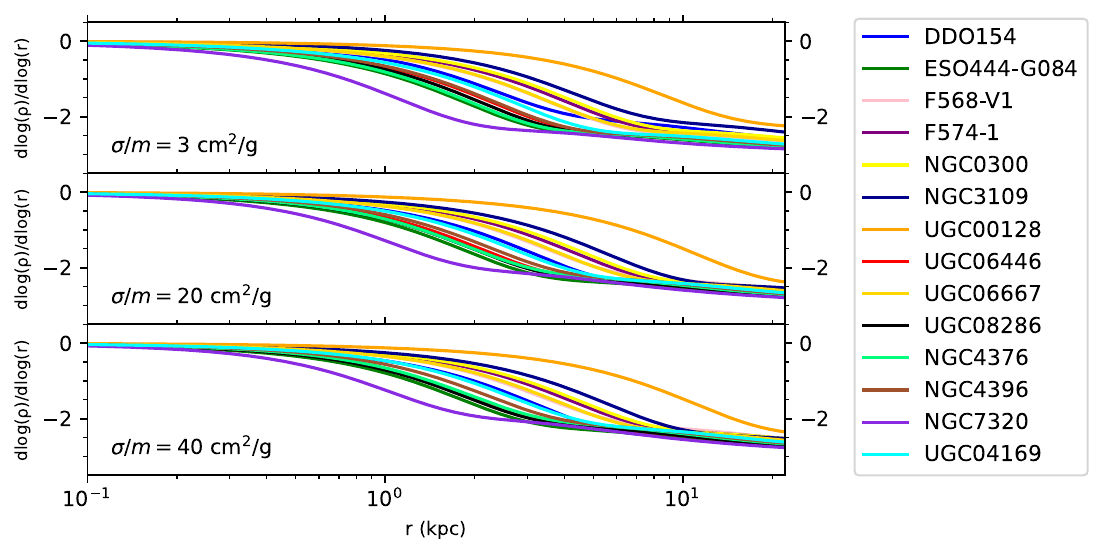}
\caption{The logarithmic slope of the density profile for all the galaxies in the sample from the best SIDM fits $\sigma/m=3~{\rm cm^2/g}$ (top), $20~{\rm cm^2/g}$ (middle), and $40~{\rm cm^2/g}$ (bottom).}
\label{fig:slope}
\end{figure*}

Fig.~\ref{fig:slope} shows radial profiles of the logarithmic density slopes from the best fits using the parametric model for all the galaxies. The profiles exhibit a overall pattern, i.e., the slope decreases from $0$ from the central region to $-2$ in the intermediate region, and then reaching $-3$ asymptotically. Since the parametric model we use is based on the Read profile, it in turn determines the overall pattern of the slope profile. We also see that the profile spans a rather wide range in the transitional intermediate region. Among the galaxies, NCG~7320 has the smallest core size, while NGC~00128 has the largest one. It is important to note that the parametric model is calibrated against the full range of gravitational evolution of an SIDM halo and the core size is not a free parameter, instead, it is dynamically determined for a given set of the initial NFW halo parameters and the cross section; see Eq.~\ref{eq:m0read}. Thus we again demonstrate the success of SIDM in explaining the diversity of the galactic rotation curves, which is reflected in the wide spread of the slope profile in the inner region as shown in Fig.~\ref{fig:slope}.

\begin{figure}[h]
\includegraphics[scale=0.6]{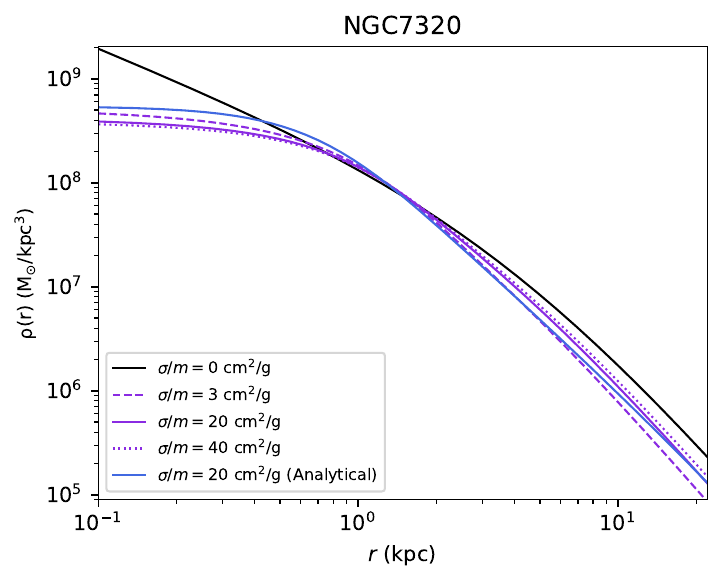}
\includegraphics[scale=0.6]{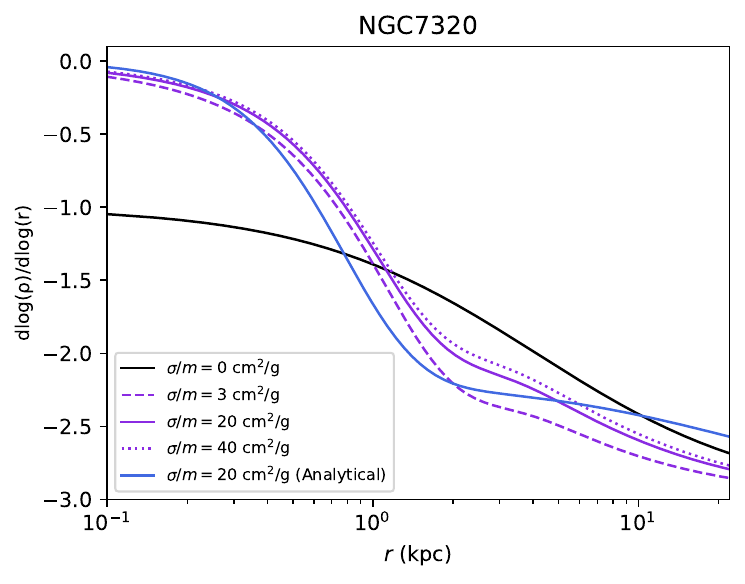}
\caption{Left panel: Density profiles of NGC7320 from the best SIDM fits $\sigma/m=0~{\rm cm^2/g}$ (solid-black), $3~{\rm cm^2/g}$ (dashed-purple), $20~{\rm cm^2/g}$ (solid-purple), and $40~{\rm cm^2/g}$ (dotted-purple). Right panel: Corresponding logarithmic slopes of the density profiles. In both panels, we also plot a profile from the best fit with $\sigma/m=20~{\rm cm^2/g}$ (solid-blue) using the analytical core-collapse density profile; see Appendix \ref{sec:app-ccp-density} for details. }
\label{fig:density}
\end{figure}

We take a close look at NGC~7320 as it has the densest halo. Fig.~\ref{fig:density} shows its best-fit density profiles using the NFW profile (black), the parametric model (magenta), as well as the analytical core-collapse profile (blue) for comparison. For the analytical profile (see Appendix~\ref{sec:app-ccp-density}), we only show the fit with $\sigma/m=20~{\rm cm^2/g}$ as a representation, as the other large cross sections yield almost the same fit. All theses fits to the rotation curve of NGC~7320 are excellent, including the one using the analytical core-collapse profile; see also Fig.~\ref{fig:example} (top). All the SIDM fits have a central core even considering a halo in the core-collapse phase and the core densities agree within $25\%$. In contrast, the inferred NFW profile has a central cusp as expected.  

Fig.~\ref{fig:density} (right) show the corresponding profiles of the logarithmic density slopes. Overall, all the SIDM fits, including the one using the analytical core-collapse profile, give rise to a similar profile, i.e., the density slope sharply drops from $0$ to $-2.2$ from the center to $r\approx2~{\rm kpc}$. In contrast, the radial change of the density slope of the NFW halo is much more mild. Thus, although both core-collapse SIDM and NFW halos can fit to a sharp rising rotation curve like NGC~7320, they are fundamentally different. For the NFW fit, the logarithmic slope approaches $-1$ for radii smaller than the scale radius $3.6~{\rm kpc}$. For the SIDM fit ($\sigma/m=40~{\rm cm^2/g}$), the halo is cored and the logarithmic slope approaches to $0$ for radii smaller than $r_c\approx0.94~{\rm kpc}$. Additionally, its scale radius reduces from its initial value $r_{s,0}\approx5.9~{\rm kpc}$ to $r_s\approx2.9~{\rm kpc}$ at $t/t_c\approx1$, following the time evolution in Eq.~\ref{eq:m0read}. It would interesting to see whether precision measurements of stellar and gas kinematics of galaxies could probe the relevant radial scale and test the subtle difference. In contrast, the density profiles from the SIDM fits with different cross sections are almost indistinguishable due to the degeneracy between cross section and concentration.

Lastly, although there is a minor difference between the slope profiles from the parametric model and the analytical core-collapse profile as indicated in Fig.~\ref{fig:density} (right), both give rise to excellent fits to the rotation curves. We also checked that the values of $\rho_c$ and $\vmax$ for all the galaxies from the fits using the analytical core-collapse profile, motivated by conducting fluid simulations, for $\sigma/m=20~\cmg$ and $40~\cmg$ agree well with the results shown in Fig.~\ref{fig:rhocvmax} based on the parametric model, which was calibrated against N-body simulations. Overall, we find good agreement between the results obtained with the parametric model and the ones derived with the analytical core-collapse
profile.

\subsection{Particle physics models}
\label{subsec:particle-physics-modeling}

To explore implications of our fits on particle physics model space of SIDM, we consider a scenario where a dark matter particle ($\chi$) couples to a light gauge boson ($\phi$) with coupling strength $g_{\chi}$~\cite{Feng:2009hw,Tulin:2017ara}. The interaction Lagrangian is given by $\mathcal{L}_{\rm{int}} = g_{\chi}\bar{\chi}\gamma^{\mu}\chi\phi_{\mu}$. For simplicity, we focus on the weakly-coupled Born limit and model $\chi\chi\rightarrow\chi\chi$ scattering with a M\o{}ller-like differential cross section~\cite{Yang:2022hkm,Girmohanta:2022dog},

\begin{equation}
    \frac{d\sigma}{d\cos\theta} = \frac{\sigma_{0}w^4 \left[\left(3\cos^2\theta + 1\right)v^4 + 4v^2w^2 + 4w^4\right]}{\left(\sin^2\theta v^4 + 4v^2w^2 + 4w^4\right)^2},
\label{eq:born-differential-sigma}
\end{equation}  
where $\sigma_0 = 4\pi\alphachi^2 / \left(\mchi^2w^4\right)$ with $\alphachi = g_{\chi}^2/4\pi$ and $w = \mphi c/\mchi$, $\mchi$ and $\mphi$ are dark matter and mediator particle masses, respectively; $v$ is the relative velocity between the two initial particles, and $\theta$ is the scattering angle. We can integrate out the angular dependence in Eq.~\ref{eq:born-differential-sigma} by introducing the viscosity cross section~\cite{Yang:2022hkm},

\begin{equation}
      \sigma_{V} = \frac{3}{2}\int d\cos\theta \sin^{2}\theta \frac{d\sigma}{d\cos\theta} = \frac{3\sigma_{0}w^8}{v^8 + 2v^6 w^2}\left[2\left(5 + 5\frac{v^2w^2}{w^4} + \frac{v^4}{w^4}\right)\ln\left(1 + \frac{v^2}{w^2}\right) - 5\left(\frac{v^4}{w^4} + 2\frac{v^2}{w^2}\right)\right].
\end{equation}

For a given halo mass, the self-scattering effect on the halo structure can be captured by the effective cross section~\cite{Outmezguine:2022bhq,Yang:2022hkm,Yang:2022zkd,Fischer:2023lvl} as we discussed in Sec.~\ref{subsec:parametric-model},

\begin{equation}
    \sigmaeff = \frac{2\int v^2 dv~d\cos\theta \frac{d\sigma}{d\cos\theta}v^5\sin^{2}\theta ~\text{exp}\left[-\frac{v^2}{4\left(\sigmavel\right)^2}\right]}{\int v^2 dv~d\cos\theta v^5\sin^{2}\theta ~\text{exp}\left[-\frac{v^2}{4\left(\sigmavel\right)^2}\right]} \approx \frac{1}{512\left(\sigmavel\right)^8}\int v^2 dv~\frac{2}{3}\sigma_{V}v^5 ~\text{exp}\left[-\frac{v^2}{4\left(\sigmavel\right)^2}\right],
\label{eq:sigmaeff}
\end{equation}
where $\sigmavel \approx0.64 \vmax$ presents the characteristic velocity dispersion of the halo.

We follow Refs.~\cite{Kaplinghat:2015aga,Andrade:2020lqq,Sagunski:2020spe} to consider constraints on the self-scattering cross section at group and cluster scales. Specifically, we adopt an upper bound of $\sigmaeffm = 0.13~\cmg$ at cluster scales, corresponding to $\vmax\approx1000~\kms$ ~\cite{Andrade:2020lqq}. Motivated by the favored range of cross sections $20\textup{--}40~{\rm cm^2/g}$ derived in this work, we also require $\sigmaeffm = (30\pm 10)~\cmg$ at a characteristic value of $\vmax \approx75~\kms$. These two conditions effectively constrain a combination of the three particle model parameters. We scan over a range of $m_\chi$ values, and for each, we use Eq.~\ref{eq:sigmaeff} and identify the allowed regions in $\alphachi$ and $\mphi/\mchi$ such that both of the following conditions are simultaneously satisfied: $\sigmaeffm = 20\textup{--}40~{\rm cm^2/g}$ at $\vmax = (75\pm 25)~\kms$ and $\sigmaeffm < 0.13\cmg$ at $\vmax = (1000 \pm 55)~\kms$.

\begin{figure}[h]
\includegraphics[scale=0.65]{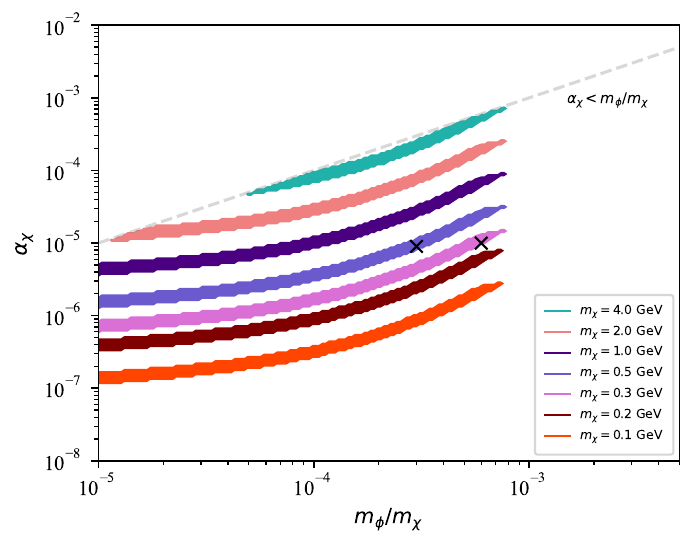}
\includegraphics[scale=0.65]{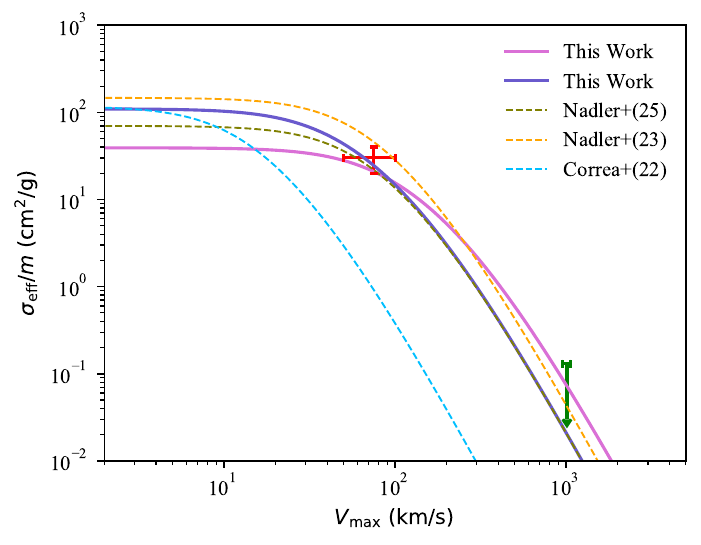}
\caption{Left panel: Favored parameter regions where the effective SIDM cross section satisfies the following conditions simultaneously: $\sigmaeffm=20\textup{--}40~{\rm cm^2/g}$ at $\vmax = (75\pm 25)~\kms$, and $\sigmaeffm < 0.13~\cmg$ at $\vmax=(1000\pm 55)~\kms$. Each of the fixed $\mchi$ values is denoted by its corresponding color as shown in the legend. The gray dashed line corresponds to the limit where the weakly-coupled Born approximation is valid, i.e., $\alphachi < \mphi/\mchi$. The black crosses denote the two examples chosen to demonstrate the velocity dependence of the cross section in the right panel. Right panel: Effective cross sections as a function of $\vmax$ for the cases $m_\chi=0.3~{\rm GeV}$ (pink) and $0.5~{\rm GeV}$ (purple). The red data point is based on our fits, and the green one is from~\cite{Andrade:2020lqq}, which is taken it to be an upper bound. For comparison, we also show the SIDM models that were simulated in Refs.~\cite{Nadler:2023nrd} (orange),~\cite{Nadler:2025jwh} (olive), and~\cite{Correa:2022dey} (blue).}
\label{fig:parameter-space}
\end{figure}

Fig.~\ref{fig:parameter-space} (left), we show seven contours in the $\alphachi\textup{--}\mphi/\mchi$ plane corresponding to dark matter particle masses in the range $\mchi = 0.1\textup{--}4~\rm{GeV}$. For the cases $m_\chi=0.3~\rm{GeV}$ and $ 0.5~\rm{GeV}$, we select two representative examples (indicated by black crosses) and show their predicted $\sigmaeffm$ as a function of $\vmax$ in Fig.~\ref{fig:parameter-space} (right). For $\mchi = 0.3~\rm{GeV}$, the model has $\sigma_{0}/m \approx 80~\cmg$ and $w \approx 180~\kms$ (pink), while for $\mchi = 0.5~\rm{GeV}$, $\sigma_{0}/m \approx 220 ~\cmg$ and $w \approx 90~\kms$ (purple). Both examples exhibit a large cross section on dwarf scales, consistent with the range favored by our fits, while satisfying the upper-limit constraints from cluster scales. Since $w=\mphi c/\mchi$, SIDM models with smaller $w$ values remain allowed, as the cluster constraints are imposed as an upper bound; see Fig.~\ref{fig:parameter-space} (left).

We further compare the SIDM models derived from our fits with those recently proposed in the literature, which feature a large self-interaction cross section in dwarf halos. Ref.~\cite{Nadler:2023nrd} performed zoom-in simulations of a group-scale halo with $V_{\rm max} \sim 350~{\rm km/s}$, using a velocity-dependent SIDM model characterized by $\sigma_0/m = 147.1~{\rm cm^2/g}$ and $w = 120~{\rm km/s}$. They demonstrated that massive subhalos with $V_{\rm max} \sim 75~{\rm km/s}$ can enter the core-collapse phase and become sufficiently dense to account for the high-density perturber observed in a strong gravitational lensing system~\cite{Vegetti:2009cz,Minor:2020hic}, which poses a challenge for the CDM model. This explanation is also expected to hold for a model with $\sigma_0/m = 70~{\rm cm^2/g}$ and $w = 120~{\rm km/s}$~\cite{Nadler:2025jwh,Yang:2024uqb}. Notably, these models have effective cross sections at $V_{\rm max} \sim 75~{\rm km/s}$ that fall within the range derived from our fits, as shown in Fig.~\ref{fig:parameter-space} (left, orange, olive). Ref.~\cite{Correa:2022dey} simulated three velocity-dependent SIDM models with $\sigma_0/m \approx 20\textup{--}100~{\rm cm^2/g}$ and $w \approx 20\textup{--}30~{\rm km/s}$. One of these models (SigmaVel100) is included in Fig.~\ref{fig:parameter-space} (left, blue) for comparison. This model was originally motivated to explain the diversity in dark matter densities observed in satellite galaxies of the Milky Way in SIDM, accommodating both core-expansion and core-collapse solutions~\cite{Correa:2020qam,Zavala:2019sjk,Nishikawa:2019lsc,Sameie:2019zfo,Kaplinghat:2019svz,Silverman:2022bhs,Slone:2021nqd}. In this model, the effective cross section remains large, $\sigma_{\rm eff}/m \gtrsim 10~{\rm cm^2/g}$, for satellite dwarf halos with $V_{\rm max} \lesssim 30~{\rm km/s}$, but drops rapidly for more massive halos, reaching $\sigma_{\rm eff}/m \approx 1~{\rm cm^2/g}$ at $V_{\rm max} \sim 75~{\rm km/s}$. It would be interesting to further investigate SIDM models that can simultaneously account for the diverse dark matter distributions across all relevant halo mass scales discussed above, including the strongly-coupled regime beyond the Born limit. We leave this for future work.

\section{Conclusions}
\label{sec:con}

In this work, we have explored the possibility of explaining the diverse galactic rotation curves in a broader range of SIDM models than discussed previously in this context. The models we consider span about one order of magnitude in cross section per mass at velocities comparable to the maximum circular velocity of the halos. We used a parametric SIDM halo model tested against N-body simulations, which takes into account the full gravothermal evolution including both core expansion and collapse phases, to fit the rotation curves of $14$ dwarf and low surface brightness galaxies. We also crosschecked our fits using an analytical core-collapse profile that builds in the temporal evolution of the core density inferred from solutions to the gravothermal fluid equations, along with the conservation of virial mass. 

The observed rotation curves were chosen to cover the full range of the diversity seen in rotation curves with $V_{\rm max}\approx 70\textup{--}80~\rm km/s$, while targeting those galaxies where the gas and stars are dynamically least important (for the range of radii where the rotation curve is measured). We also picked a few rotation curves with larger and smaller $V_{\rm max}$ values to provide a comparison. We found that SIDM can provide excellent fits (Fig.~\ref{fig:combo}) to both rotation curves that rise rapidly and very gently (Fig.~\ref{fig:example}). Compared to the SIDM fits with $\sigma/m=3~{\rm cm^2/g}$, which have been explored in a lot of detail, those with $\sigma/m=20\textup{--}40~{\rm cm^2/g}$ require a narrower spread in the halo concentration (Fig.~\ref{fig:rhocdc200}). This is because a larger cross section further speeds up the gravothermal evolution of the halo and amplifies the diversity of the halo density encoded in the scatter of the concentration-mass relation. For $\sigma/m$ taking on values around $20\textup{--}40~{\rm cm^2/g}$, the galactic halos with larger concentrations are further into the core-collapse phase within the Hubble time.

Our SIDM fits demonstrate a unified scenario of understanding the halo properties of the history of gravothermal evolution by placing each galaxy on the universal evolution trajectory of the halo (Fig.~\ref{fig:rhotau}). A galaxy with a relatively higher inner density indicates that its halo is at later stage of the evolution, and vice versa; however, as mentioned previously, if a galaxy undergoes a major merger it may reset the core-collapse timescale. Furthermore, all our fits result in similar distributions in the halo parameters, i.e., $\rho_c\textup{--}V_{\rm max}$ (Fig.~\ref{fig:rhocvmax}) and $\rho_c\textup{--}r_c$ (Fig.~\ref{fig:rhocrc}), indicating there is degeneracy in fitting the rotation curves as expected. We also discussed logarithmic density slopes of the SIDM halos (Fig.~\ref{fig:slope}) and showed that it is difficult to distinguish between the density profiles halos in the core expansion or core-collapse phases (Fig.~\ref{fig:density}) purely based on the rotation curves. 
We also demonstrated that the fits using an analytical core-collapse profile agree with those using the parametric model. Lastly, we explored the implications of our fits for the particle physics parameter space of SIDM by incorporating constraints from cluster scales and comparing the favored models with those recently proposed in the literature (Fig.~\ref{fig:parameter-space}).

More work with N-body simulations is required to validate the simple models that we have used to fit the rotation curves and infer the range of cross section values for explaining the diversity of rotation curves. The key point we would like to emphasize is that if the dark matter halos are predominantly in the core-collapse phase, then SIDM models have an in-built mechanism to explain both the slowly rising rotation curves as well as the sharply rising ones. Note that the core-collapse phase is very long and the central density only increases mildly for the majority of this evolution; hence the rotation curves in this phase are very similar to those in the core expansion but the required halo concentrations are closer to the median values. However, if the cross section gets very large, then many galaxies will be in a core-collapse regime where the core radius is very small and the rotation curves are sharply rising like that of NGC 7320. The paucity of such sharply rising rotation curves in baryon-poor galaxies can be used to put an upper limit on the cross section value at the relevant velocity scales.

Overall, our results show that SIDM models with $\sigma/m$ around $20~{\rm cm^2/g}$ or larger can address the diversity ``problem" of dark-matter dominated spiral galaxies better than the model space with $\sigma/m\sim{\cal O}(1)~{\rm cm^2/g}$, because with larger cross sections there are many fewer outliers in concentration than with smaller cross sections. Intriguingly, such large cross section values are also motivated by observations of diversity in the dark matter densities of the satellite galaxies of the Milky Way~\cite{Valli:2017ktb,Kaplinghat:2019svz,Silverman:2022bhs,Yang:2022mxl,Zhang:2024ggu} and gas-rich ultra-diffuse  galaxies in the field~\cite{ManceraPina:2019zih,PinaMancera:2021wpc,Kong:2022oyk,ManceraPina:2024ybj}, as well as inferences of dense substructures discovered in strong lensing~\cite{Vegetti:2009cz,Minor:2020hic,Nadler:2023nrd,Sengul:2022edu,Zhang:2023wda,Meneghetti:2020yif,Yang:2021kdf,Dutra:2024qac,Gilman:2021sdr} and stellar stream~\cite{Zhang:2024fib} systems. It would be interesting to explore particle physics realizations of SIDM that can simultaneously explain these observations.

\begin{acknowledgments}
We thank Daneng Yang for helpful discussion. MK acknowledges support by the National Science Foundation under grant No. PHY-2210283. MV acknowledges support from the project ``Theoretical Particle Physics and Cosmology (TPPC)'' funded by INFN. HBY acknowledges support by the Department of Energy under grant No. de-sc0008541 and the John Templeton Foundation under grant ID \#61884. The opinions expressed in this publication are those of the authors and do not necessarily reflect the views of the John Templeton Foundation.

\end{acknowledgments}

\bibliographystyle{apsrev4-1}
\bibliography{references}

\newpage

\appendix

\section{An analytical core-collapse density profile}

\label{sec:app-ccp-density}

Towards the end of Sec.~\ref{subsec:parametric-model}, we discussed three conditions for obtaining an analytical core-collapse density profile. Motivated by Ref.~\cite{Jiang:2021foz}, we consider the following density profile based on the conducting fluid model

\begin{equation}
    \rho(x | x_{c}, b, g) = \frac{g \rho_{s,0} \tanh^\alpha\left(x/x_c\right)}
    {x^{\alpha}\left(b + x\right)^{3-\alpha}} = g \rho_{s,0} \Tilde{\rho}(x, x_c, b),
\label{eq:main density profile}
\end{equation}
where $\alpha = 2.19$, $x_{c} = r_{c}/r_{s,0}$, $x = r/r_{s,0}$, and $\Tilde{\rho}(x, x_c, b) = \frac{\tanh^\alpha\left(x/x_c\right)}{x^{\alpha}\left(b + x\right)^{3-\alpha}}$. 

We have introduced a normalization parameter $g$, and a parameter $b$, which will control the transition between the different regions of the slope of the density profile. The mass profile is given by $M(x|x_c,b,g) = 4 \pi \rho_{s,0} r_{s,0}^3  \Tilde{M}(x,x_c,b,g)$ with,
\begin{equation}
    \Tilde{M}(x, x_{c}, b, g) = g \int_{0}^{x} dx ~\Tilde{\rho}(x, x_{c}, b)~x^{2}.
\label{eq:mass-profile}
\end{equation}

We first determine $g$ by demanding that the enclosed mass within $r_{200}$ of Eq.~\ref{eq:main density profile} is equal to that of the initial NFW $M_{200}$

\begin{equation}
    g = \frac{\ln\left(1+c_{200}\right) - \frac{c_{200}}{1 + c_{200}}}{\int_{0}^{c_{200}} dx ~\Tilde{\rho}(x, x_{c}, b)~x^{2}}.
\label{eq:g-constraint}
\end{equation}

In order to determine $b$, we express the central density of Eq.~\ref{eq:main density profile} as,
\begin{equation}\label{eq:rhoc-rhos-calc}
    \frac{\rho(0)}{\rho_{s,0}} = \frac{g}{x_{c}^{\alpha}b^{(3-\alpha)}},
\end{equation}

\noindent where we have taken the $r\rightarrow 0$ limit. This ratio is a universal quantity for an SIDM halo and is determined by the gravothermal evolution trajectory and is related to the time evolution. In practice, we use the ratio of the central 1D velocity dispersion, $v(0)$, of dark matter particles to $\vmax$, $v(0)/\vmax$, as a proxy for the time evolution as used in \cite{Outmezguine:2022bhq}, see the left panel of their Fig. 2. For a given halo, this ratio can be determined by solving the Jeans equation:

\begin{equation}
    (v(0)/\vmax)^2 = x_{c}^{\alpha} ~b^{3-\alpha} ~2.15^{2} \int_{0}^{\infty} \Tilde{M}(x, x_{c}, b, g) \Tilde{\rho}(x, x_{c}, b) x^{-2} dx.
\label{eq:vc-vmax}
\end{equation}

Ref.~\cite{Outmezguine:2022bhq} solved solved the gravothermal equations for a fluid model and have determined a universal relation for the central density evolution as a function of the velocity dispersion: $\rho(0)/\rho_{s,0}~\textup{--}~v(0)/\vmax$. For each $v(0)/\vmax$ we determine $b$ by imposing the following condition, 

\begin{equation}\label{eq:enforce-theory-curve}
    \frac{\rho(0)}{\rho_{s,0}} = \frac{g}{x_{c}^{\alpha}b^{(3-\alpha)}} = \left(\frac{\rho(0)}{\rho_{s,0}}\right)_{\text{theory}},
\end{equation}

\noindent where $\left(\rho(0)/\rho_{s,0}\right)_{\text{theory}}$ at $v(0)/\vmax$ is taken from the curve ($n = 0$) in the left panel of Fig.~2~\cite{Outmezguine:2022bhq}. Note that both $g$ and $b$ only depend on $c_{200}$ for a given value of $x_{c}$. Therefore, we solve Eqs.~\ref{eq:g-constraint} and~\ref{eq:enforce-theory-curve} simultaneously as functions of $c_{200}$ and $x_{c}$, which result from the core radius $r_{c}$, and the initial NFW parameters $r_{s,0}$ and $\rho_{s,0}$. The value of $b$ required to get the right $v(0)/\vmax$ turns out to be in the range from $3$ to $4$, and the value of $g$ is close to $1.03$ for most values of $x_{c}$ and $c_{200}$. One the values of $b$ and $g$ are known, we can calculate the cross section by utilizing Eq.~\ref{eq:rhoc-rhos-calc}, Eq.~\ref{eq:tc0}, and the curve ($n = 0$) in the right panel of Fig. 1 \cite{Outmezguine:2022bhq}.

\section{Contour plots}
\label{sec:app-contour-plots}

\begin{figure*}[h]
\includegraphics[scale=0.8]{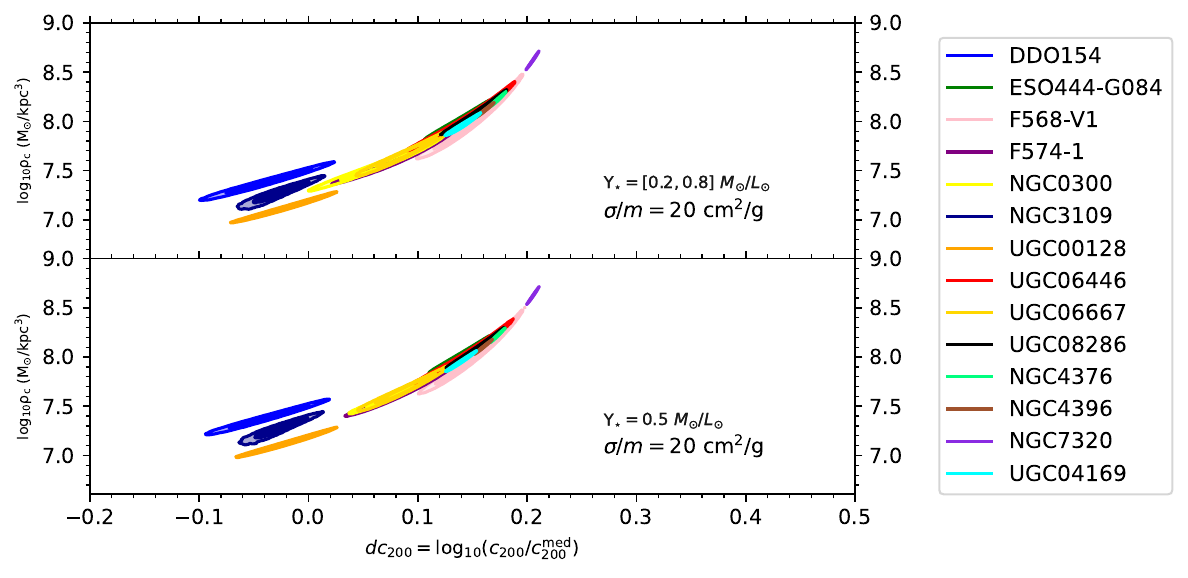}
\caption{The $1\sigma$ and $2\sigma$ contours for the core density vs concentration normalized to the cosmological median from the SIDM fits $20~{\rm cm^2/g}$ with varying $\Upsilon_{\star}$ (top) and $20~{\rm cm^2/g}$ (bottom) with fixed $\Upsilon_{\star}$. The $1\sigma$ scatter of the concentration is $0.11$ dex.}
\label{fig:rhocdc200_contours}
\end{figure*}

In Fig.~\ref{fig:rhocdc200_contours}, we show the $1\sigma$ and $2\sigma$ contours in the $\log_{10}\rho_c\textup{--}dc_{200}$ plane for $\sigma/m=20~{\rm cm^2/g}$ assuming the varying mass-to-light ratio in the range $0.2\textup{--0.8}~M_\odot/L_\odot$ (top) and the fixed one $0.5~M_\odot/L_\odot$ (bottom). We note that the core density and the concentration ratio are correlated, while being tightly constrained regardless of the varying stellar mass-to-light ratio.

\begin{figure}[h]
\includegraphics[scale=0.8]{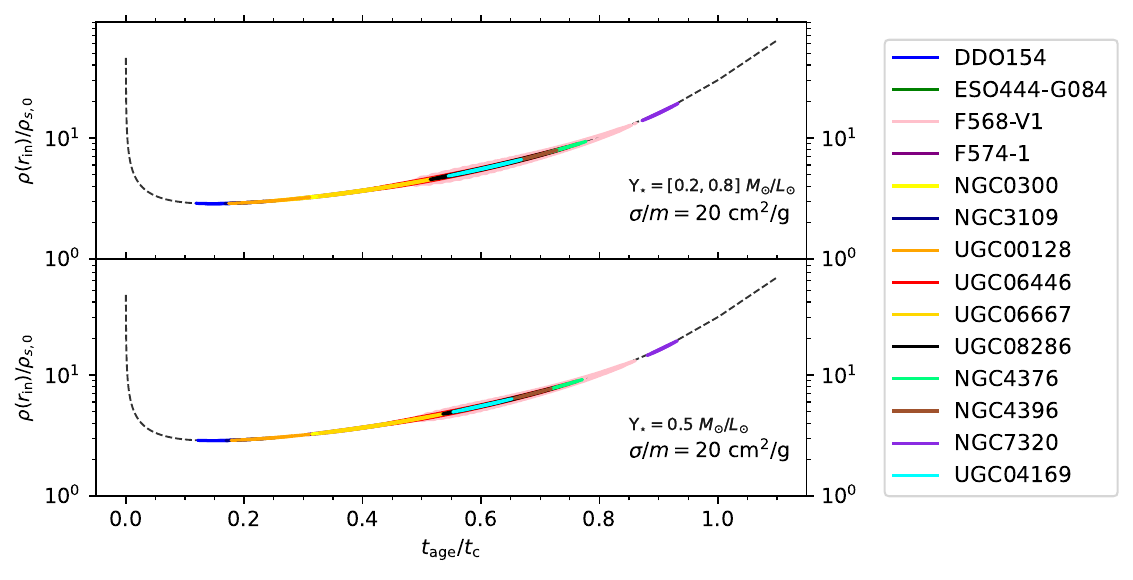}
\caption{The $1\sigma$ and $2\sigma$ contours for the central density vs gravothermal time from the SIDM fits $20~{\rm cm^2/g}$ with varying $\Upsilon_{\star}$ (top) and $20~{\rm cm^2/g}$ with fixed $\Upsilon_{\star}$ (bottom). The central density, evaluated at the radius $r_{\text{in}} = 10^{-3}r_{s,0}$, is normalized to the scale radius of its corresponding initial NFW halo, and $t_{\rm age}=10~{\rm Gyr}$, and $t_c$ is calculated using Eq.~\ref{eq:tc0}. The dashed curve denotes the universal evolution trajectory of an SIDM halo characterized by Eq.~\ref{eq:m0read}. }
\label{fig:rhotau_contours}
\end{figure}

\begin{figure}[h]
\includegraphics[scale=0.8]{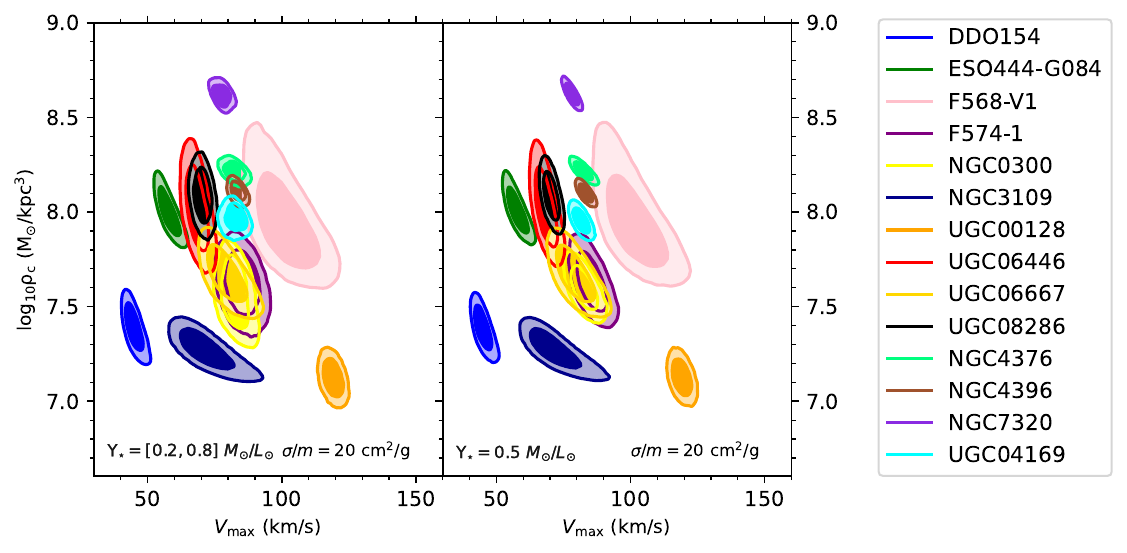}
\caption{The $1\sigma$ and $2\sigma$ contours for the core density vs maximum halo circular velocity for the SIDM fits with $20~{\rm cm^2/g}$ with varying $\Upsilon_{\star}$ (left) and $20~{\rm cm^2/g}$ with fixed $\Upsilon_{\star}$ (right).}
\label{fig:rhocvmax_contours}
\end{figure}

\begin{figure}[h]
\includegraphics[scale=0.8]{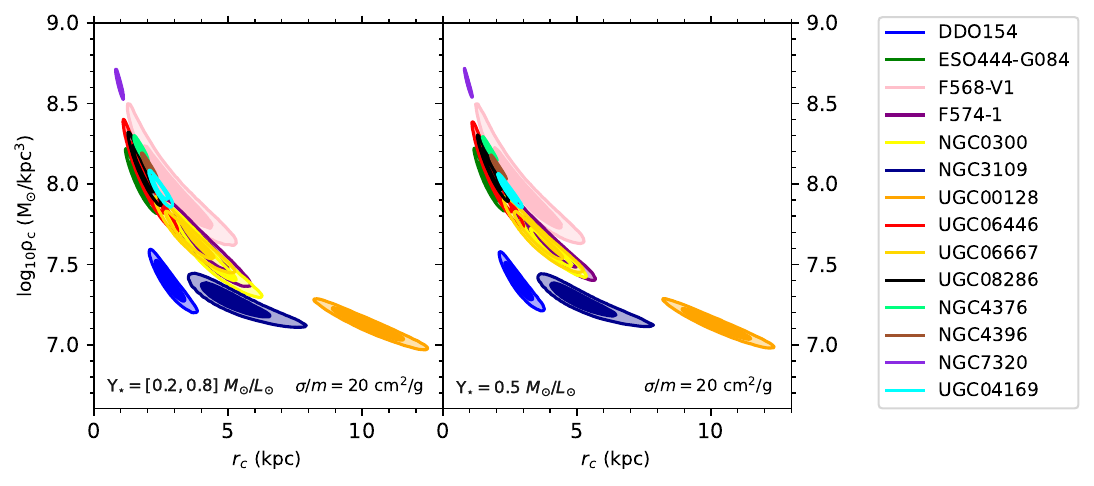}
\caption{The $1\sigma$ and $2\sigma$ contours for the core density vs core radius for the SIDM fits with $20~{\rm cm^2/g}$ with varying $\Upsilon_{\star}$ (left) and $20~{\rm cm^2/g}$ with fixed $\Upsilon_{\star}$ (right).}
\label{fig:rhocrc_contours}
\end{figure}

Comparing the other corresponding plots, i.e., Fig.~\ref{fig:rhotau} with Fig.~\ref{fig:rhotau_contours}, Fig.~\ref{fig:rhocvmax} with Fig.~\ref{fig:rhocvmax_contours}, and Fig.~\ref{fig:rhocrc} with Fig.~\ref{fig:rhocrc_contours}, shows the same pattern, i.e., the parameters are correlated with each other but are robust to the variation of the mass-to-light ratio over the range $0.2\textup{--}0.8~M_{\odot}/L_{\odot}$. This result applies to the fits with the other cross sections as well.

\section{Summary of parameters inferred from the SIDM fits}
\label{sec:app-parameters-table}

In Table~\ref{tab:numerics1} and Table~\ref{tab:numerics2}, we list values of the key parameters for each galaxy from the SIDM fits based on the the parametric model in Sec.~\ref{subsec:parametric-model}. Each table is divided horizontally into groups of 3, the rows in each group corresponds to $3~\cmg$, $20~\cmg$, and $40~\cmg$, respectively. The results shown are for the fixed mass-to-light ratio $\Upsilon_{\star,\rm{disk}}  = 0.5~ M_{\odot}/L_{\odot}$ fits for each cross section. 

\begin{table}[h]
    \centering
    \begin{tabular}{|c|c|c|c|c|c|c|c|c|}
        \hline
         Galaxy & $\vmax~(\kms)$ & $\rmax~(\kpc)$ & $\log_{10}\rho_{s,0}~(M_{\odot}/\kpc^{3})$ & $r_{s,0}~(\kpc)$ & $\log_{10}M_{200}~(M_{\odot})$ & $c_{200}$\\\hline
 DDO154      & $46.91^{+2.9}_{-2.42}$   & $12.69^{+2.63}_{-2.11}$ & $6.74^{+0.12}_{-0.12}$ & $5.87^{+1.22}_{-1.02}$  & $10.31^{+0.11}_{-0.1}$  & $9.83^{+1.19}_{-1.09}$  \\
 DDO154      & $44.87^{+2.27}_{-1.91}$  & $10.13^{+1.38}_{-1.14}$ & $6.92^{+0.07}_{-0.08}$ & $4.54^{+0.62}_{-0.53}$  & $10.2^{+0.08}_{-0.08}$  & $11.68^{+0.83}_{-0.82}$ \\
 DDO154      & $44.73^{+2.27}_{-1.96}$  & $10.81^{+1.33}_{-1.11}$ & $6.86^{+0.06}_{-0.06}$ & $4.88^{+0.59}_{-0.51}$  & $10.22^{+0.08}_{-0.07}$ & $10.99^{+0.62}_{-0.63}$ \\\hline
 ESO444-G084 & $59.39^{+2.51}_{-2.16}$  & $6.14^{+0.77}_{-0.67}$  & $7.6^{+0.07}_{-0.07}$  & $2.75^{+0.35}_{-0.31}$  & $10.35^{+0.07}_{-0.07}$ & $21.64^{+1.48}_{-1.4}$  \\
 ESO444-G084 & $57.69^{+2.67}_{-2.39}$  & $8.07^{+0.81}_{-0.72}$  & $7.31^{+0.04}_{-0.05}$ & $3.72^{+0.37}_{-0.34}$  & $10.41^{+0.07}_{-0.07}$ & $16.73^{+0.7}_{-0.69}$  \\
 ESO444-G084 & $56.99^{+2.68}_{-2.35}$  & $9.9^{+0.92}_{-0.81}$   & $7.12^{+0.04}_{-0.04}$ & $4.58^{+0.43}_{-0.39}$  & $10.45^{+0.07}_{-0.07}$ & $14.06^{+0.51}_{-0.52}$ \\\hline
 F568-V1     & $105.32^{+7.23}_{-6.45}$ & $12.05^{+2.46}_{-2.14}$ & $7.51^{+0.12}_{-0.12}$ & $5.41^{+1.09}_{-0.98}$  & $11.13^{+0.12}_{-0.12}$ & $20.0^{+2.35}_{-2.07}$  \\
 F568-V1     & $100.9^{+7.74}_{-6.84}$  & $17.2^{+2.78}_{-2.45}$  & $7.14^{+0.08}_{-0.07}$ & $7.94^{+1.27}_{-1.17}$  & $11.19^{+0.11}_{-0.12}$ & $14.29^{+1.06}_{-0.93}$ \\
 F568-V1     & $99.92^{+7.92}_{-7.58}$  & $21.71^{+3.3}_{-3.14}$  & $6.93^{+0.07}_{-0.06}$ & $10.04^{+1.53}_{-1.51}$ & $11.24^{+0.11}_{-0.13}$ & $11.75^{+0.81}_{-0.67}$ \\\hline
 F574-1      & $87.13^{+3.77}_{-3.4}$   & $14.0^{+2.48}_{-1.99}$  & $7.21^{+0.12}_{-0.13}$ & $6.33^{+1.16}_{-0.96}$  & $10.98^{+0.08}_{-0.08}$ & $15.21^{+1.72}_{-1.68}$ \\
 F574-1      & $85.34^{+3.83}_{-3.5}$   & $16.53^{+2.01}_{-1.7}$  & $7.04^{+0.06}_{-0.07}$ & $7.56^{+0.88}_{-0.78}$  & $11.01^{+0.07}_{-0.07}$ & $13.0^{+0.77}_{-0.8}$   \\
 F574-1      & $83.95^{+3.94}_{-3.59}$  & $19.44^{+2.11}_{-1.78}$ & $6.87^{+0.05}_{-0.06}$ & $8.96^{+0.96}_{-0.84}$  & $11.03^{+0.07}_{-0.07}$ & $11.18^{+0.54}_{-0.56}$ \\\hline
 NGC0300     & $84.17^{+3.46}_{-3.07}$  & $14.23^{+2.03}_{-1.67}$ & $7.16^{+0.09}_{-0.1}$  & $6.45^{+0.95}_{-0.81}$  & $10.95^{+0.07}_{-0.07}$ & $14.59^{+1.26}_{-1.27}$ \\
 NGC0300     & $82.61^{+3.4}_{-3.14}$   & $16.27^{+1.63}_{-1.44}$ & $7.03^{+0.05}_{-0.05}$ & $7.42^{+0.72}_{-0.66}$  & $10.97^{+0.06}_{-0.06}$ & $12.85^{+0.61}_{-0.62}$ \\
 NGC0300     & $81.29^{+3.47}_{-3.24}$  & $18.96^{+1.72}_{-1.53}$ & $6.87^{+0.04}_{-0.04}$ & $8.74^{+0.78}_{-0.72}$  & $10.99^{+0.06}_{-0.06}$ & $11.12^{+0.43}_{-0.45}$ \\\hline
 NGC3109     & $73.88^{+7.25}_{-5.39}$  & $21.59^{+4.44}_{-3.41}$ & $6.67^{+0.09}_{-0.09}$ & $9.97^{+2.05}_{-1.63}$  & $10.92^{+0.14}_{-0.12}$ & $9.23^{+0.83}_{-0.78}$  \\
 NGC3109     & $71.08^{+7.06}_{-5.02}$  & $18.43^{+3.32}_{-2.42}$ & $6.8^{+0.06}_{-0.07}$  & $8.27^{+1.48}_{-1.12}$  & $10.84^{+0.14}_{-0.11}$ & $10.43^{+0.63}_{-0.65}$ \\
 NGC3109     & $70.81^{+7.05}_{-5.02}$  & $19.98^{+3.47}_{-2.49}$ & $6.72^{+0.05}_{-0.06}$ & $9.06^{+1.56}_{-1.16}$  & $10.85^{+0.14}_{-0.11}$ & $9.65^{+0.5}_{-0.54}$   \\\hline
 UGC00128    & $120.81^{+2.68}_{-2.49}$ & $38.84^{+4.85}_{-4.16}$ & $6.59^{+0.09}_{-0.1}$  & $17.91^{+2.27}_{-2.03}$ & $11.58^{+0.05}_{-0.04}$ & $8.54^{+0.77}_{-0.75}$  \\
 UGC00128    & $119.11^{+2.41}_{-2.34}$ & $36.93^{+2.9}_{-2.62}$  & $6.64^{+0.05}_{-0.06}$ & $16.6^{+1.28}_{-1.19}$  & $11.55^{+0.04}_{-0.04}$ & $9.0^{+0.46}_{-0.47}$   \\
 UGC00128    & $117.89^{+2.51}_{-2.43}$ & $40.25^{+2.8}_{-2.54}$  & $6.55^{+0.04}_{-0.05}$ & $18.33^{+1.22}_{-1.15}$ & $11.56^{+0.03}_{-0.04}$ & $8.22^{+0.33}_{-0.35}$  \\\hline
 UGC06446    & $71.4^{+2.52}_{-2.36}$   & $7.56^{+1.23}_{-1.06}$  & $7.58^{+0.11}_{-0.12}$ & $3.39^{+0.56}_{-0.48}$  & $10.6^{+0.07}_{-0.07}$  & $21.25^{+2.27}_{-2.16}$ \\
 UGC06446    & $68.3^{+2.84}_{-2.57}$   & $9.94^{+1.17}_{-0.98}$  & $7.28^{+0.06}_{-0.07}$ & $4.59^{+0.53}_{-0.46}$  & $10.64^{+0.07}_{-0.07}$ & $16.2^{+0.93}_{-0.96}$  \\
 UGC06446    & $67.15^{+2.77}_{-2.58}$  & $12.25^{+1.19}_{-1.06}$ & $7.08^{+0.05}_{-0.05}$ & $5.66^{+0.55}_{-0.51}$  & $10.68^{+0.07}_{-0.07}$ & $13.53^{+0.62}_{-0.64}$ \\\hline
 UGC06667    & $80.51^{+5.05}_{-4.22}$  & $12.34^{+2.38}_{-1.86}$ & $7.25^{+0.11}_{-0.12}$ & $5.58^{+1.1}_{-0.89}$   & $10.86^{+0.11}_{-0.1}$  & $15.81^{+1.69}_{-1.65}$ \\
 UGC06667    & $79.1^{+5.09}_{-4.33}$   & $14.68^{+2.12}_{-1.72}$ & $7.07^{+0.06}_{-0.07}$ & $6.71^{+0.95}_{-0.8}$   & $10.9^{+0.1}_{-0.09}$   & $13.45^{+0.8}_{-0.83}$  \\
 UGC06667    & $78.16^{+5.14}_{-4.43}$  & $17.45^{+2.29}_{-1.88}$ & $6.91^{+0.05}_{-0.06}$ & $8.05^{+1.04}_{-0.89}$  & $10.93^{+0.1}_{-0.09}$  & $11.51^{+0.58}_{-0.59}$ \\\hline
 UGC08286    & $73.55^{+1.82}_{-1.75}$  & $7.06^{+0.72}_{-0.64}$  & $7.66^{+0.07}_{-0.07}$ & $3.17^{+0.32}_{-0.29}$  & $10.61^{+0.05}_{-0.05}$ & $22.9^{+1.46}_{-1.47}$  \\
 UGC08286    & $70.45^{+1.98}_{-1.89}$  & $10.04^{+0.7}_{-0.62}$  & $7.29^{+0.04}_{-0.04}$ & $4.64^{+0.32}_{-0.29}$  & $10.68^{+0.05}_{-0.05}$ & $16.45^{+0.55}_{-0.57}$ \\
 UGC08286    & $69.74^{+2.01}_{-1.94}$  & $12.69^{+0.78}_{-0.72}$ & $7.08^{+0.03}_{-0.03}$ & $5.87^{+0.36}_{-0.35}$  & $10.73^{+0.04}_{-0.05}$ & $13.55^{+0.36}_{-0.39}$ \\\hline
 NGC4376     & $83.36^{+2.29}_{-2.12}$  & $6.88^{+0.38}_{-0.35}$  & $7.79^{+0.02}_{-0.02}$ & $3.12^{+0.17}_{-0.16}$  & $10.73^{+0.04}_{-0.04}$ & $25.54^{+0.54}_{-0.56}$ \\
 NGC4376     & $82.29^{+2.31}_{-2.16}$  & $11.73^{+0.56}_{-0.51}$ & $7.29^{+0.02}_{-0.02}$ & $5.42^{+0.26}_{-0.25}$  & $10.88^{+0.04}_{-0.04}$ & $16.44^{+0.25}_{-0.26}$ \\
 NGC4376     & $83.34^{+2.48}_{-2.31}$  & $15.74^{+0.76}_{-0.71}$ & $7.05^{+0.02}_{-0.02}$ & $7.28^{+0.35}_{-0.34}$  & $10.97^{+0.04}_{-0.04}$ & $13.15^{+0.19}_{-0.2}$  \\\hline
 NGC4396     & $85.96^{+1.7}_{-1.58}$   & $8.08^{+0.38}_{-0.36}$  & $7.68^{+0.03}_{-0.03}$ & $3.64^{+0.17}_{-0.17}$  & $10.81^{+0.03}_{-0.03}$ & $23.22^{+0.53}_{-0.53}$ \\
 NGC4396     & $83.23^{+1.71}_{-1.62}$  & $12.43^{+0.47}_{-0.44}$ & $7.25^{+0.02}_{-0.02}$ & $5.74^{+0.22}_{-0.21}$  & $10.91^{+0.03}_{-0.03}$ & $15.85^{+0.22}_{-0.23}$ \\
 NGC4396     & $83.55^{+1.77}_{-1.7}$   & $16.22^{+0.59}_{-0.57}$ & $7.03^{+0.01}_{-0.01}$ & $7.5^{+0.27}_{-0.28}$   & $10.98^{+0.03}_{-0.03}$ & $12.86^{+0.17}_{-0.17}$ \\\hline
 NGC7320     & $79.54^{+1.37}_{-1.32}$  & $5.0^{+0.18}_{-0.18}$   & $8.01^{+0.02}_{-0.02}$ & $2.3^{+0.08}_{-0.08}$   & $10.59^{+0.03}_{-0.03}$ & $31.08^{+0.46}_{-0.48}$ \\
 NGC7320     & $78.14^{+1.65}_{-1.6}$   & $9.84^{+0.36}_{-0.35}$  & $7.4^{+0.01}_{-0.01}$  & $4.55^{+0.16}_{-0.17}$  & $10.78^{+0.03}_{-0.03}$ & $18.13^{+0.22}_{-0.22}$ \\
 NGC7320     & $76.25^{+1.87}_{-1.82}$  & $12.84^{+0.52}_{-0.51}$ & $7.15^{+0.01}_{-0.01}$ & $5.94^{+0.24}_{-0.24}$  & $10.82^{+0.04}_{-0.04}$ & $14.41^{+0.19}_{-0.19}$ \\\hline
 UGC04169    & $83.82^{+1.99}_{-1.88}$  & $9.03^{+0.6}_{-0.55}$   & $7.56^{+0.04}_{-0.04}$ & $4.05^{+0.27}_{-0.26}$  & $10.82^{+0.04}_{-0.04}$ & $20.96^{+0.77}_{-0.78}$ \\
 UGC04169    & $81.58^{+2.09}_{-1.98}$  & $12.89^{+0.65}_{-0.6}$  & $7.21^{+0.02}_{-0.02}$ & $5.95^{+0.3}_{-0.29}$   & $10.9^{+0.04}_{-0.04}$  & $15.17^{+0.31}_{-0.32}$ \\
 UGC04169    & $81.58^{+2.1}_{-2.03}$   & $16.45^{+0.76}_{-0.73}$ & $6.99^{+0.02}_{-0.02}$ & $7.61^{+0.35}_{-0.35}$  & $10.96^{+0.04}_{-0.04}$ & $12.48^{+0.22}_{-0.22}$ \\
\hline
\end{tabular}
    \caption{From the left to right columns: galaxy name, the maximal circular velocity ($V_{\rm max}$) and its associated radius ($R_{\rm max}$) of the fitted SIDM halo, the scale density ($\rho_{s,0}$) and radius ($r_{r,0}$) of the initial NFW halo, the total halo mass ($M_{200}$) and concentration ($c_{200}$). For each galaxy, the first, second, and third rows are for $\sigmam=3~\cmg$, $20~\cmg$, and $40~\cmg$, respectively.}
    \label{tab:numerics1}
\end{table}{}

\begin{table}[h]
    \centering
    \begin{tabular}{|c|c|c|c|c|c|c|}
    \hline
    Galaxy & $r_{c}/r_{s,0}$ & $r_{c}~(\kpc)$ & $\rho_{c}/\rho_{s,0}$ & $\log_{10}\rho_{c}~(M_{\odot}/\kpc^{3})$ & $\chi^{2}/\text{d.o.f.}$ & $\text{min}\left(\chi^{2}/\text{d.o.f.}\right)$\\\hline

 DDO154      & $0.36^{+0.03}_{-0.03}$ & $2.09^{+0.24}_{-0.21}$  & $3.97^{+0.39}_{-0.36}$ & $7.4^{+0.09}_{-0.08}$  & $11.21^{+18.42}_{-7.44}$ & 1.2208 \\
 DDO154      & $0.62^{+0.0}_{-0.0}$   & $2.83^{+0.37}_{-0.33}$  & $2.1^{+0.03}_{-0.02}$  & $7.38^{+0.08}_{-0.08}$ & $10.23^{+17.17}_{-7.0}$  & 1.2062 \\
 DDO154      & $0.6^{+0.01}_{-0.02}$  & $2.91^{+0.41}_{-0.36}$  & $2.11^{+0.03}_{-0.02}$ & $7.36^{+0.08}_{-0.07}$ & $10.31^{+17.5}_{-7.04}$  & 1.2208 \\\hline
 ESO444-G084 & $0.62^{+0.0}_{-0.0}$   & $1.72^{+0.2}_{-0.19}$   & $2.11^{+0.03}_{-0.02}$ & $8.06^{+0.08}_{-0.07}$ & $2.59^{+0.69}_{-0.34}$   & 2.0026 \\
 ESO444-G084 & $0.45^{+0.02}_{-0.03}$ & $1.69^{+0.25}_{-0.24}$  & $2.51^{+0.1}_{-0.08}$  & $8.0^{+0.08}_{-0.08}$  & $2.56^{+0.68}_{-0.34}$   & 2.1056 \\
 ESO444-G084 & $0.35^{+0.02}_{-0.03}$ & $1.62^{+0.26}_{-0.25}$  & $2.95^{+0.15}_{-0.12}$ & $7.98^{+0.09}_{-0.08}$ & $2.45^{+0.68}_{-0.33}$   & 2.0026 \\\hline
 F568-V1     & $0.62^{+0.01}_{-0.02}$ & $3.33^{+0.72}_{-0.7}$   & $2.09^{+0.04}_{-0.01}$ & $7.99^{+0.15}_{-0.13}$ & $0.28^{+0.35}_{-0.17}$   & 0.0905 \\
 F568-V1     & $0.39^{+0.04}_{-0.06}$ & $3.11^{+0.85}_{-0.85}$  & $2.77^{+0.3}_{-0.19}$  & $7.93^{+0.18}_{-0.14}$ & $0.3^{+0.35}_{-0.18}$    & 0.0698 \\
 F568-V1     & $0.29^{+0.04}_{-0.06}$ & $2.96^{+0.88}_{-0.92}$  & $3.31^{+0.42}_{-0.27}$ & $7.91^{+0.21}_{-0.15}$ & $0.32^{+0.34}_{-0.17}$   & 0.0905 \\\hline
 F574-1      & $0.59^{+0.02}_{-0.03}$ & $3.73^{+0.49}_{-0.45}$  & $2.28^{+0.14}_{-0.1}$  & $7.68^{+0.11}_{-0.11}$ & $0.45^{+0.54}_{-0.27}$   & 0.1059 \\
 F574-1      & $0.53^{+0.03}_{-0.03}$ & $3.99^{+0.68}_{-0.62}$  & $2.28^{+0.1}_{-0.08}$  & $7.62^{+0.11}_{-0.1}$  & $0.45^{+0.54}_{-0.26}$   & 0.096  \\
 F574-1      & $0.43^{+0.03}_{-0.04}$ & $3.88^{+0.7}_{-0.64}$   & $2.59^{+0.15}_{-0.12}$ & $7.6^{+0.11}_{-0.1}$   & $0.46^{+0.53}_{-0.27}$   & 0.1059 \\\hline
 NGC0300     & $0.57^{+0.02}_{-0.03}$ & $3.7^{+0.39}_{-0.36}$   & $2.35^{+0.12}_{-0.1}$  & $7.64^{+0.08}_{-0.08}$ & $1.01^{+0.6}_{-0.29}$    & 0.5873 \\
 NGC0300     & $0.54^{+0.02}_{-0.02}$ & $4.02^{+0.53}_{-0.5}$   & $2.24^{+0.06}_{-0.05}$ & $7.59^{+0.08}_{-0.08}$ & $0.99^{+0.61}_{-0.29}$   & 0.6135 \\
 NGC0300     & $0.45^{+0.02}_{-0.03}$ & $3.93^{+0.55}_{-0.51}$  & $2.53^{+0.1}_{-0.08}$  & $7.57^{+0.08}_{-0.08}$ & $0.97^{+0.6}_{-0.29}$    & 0.5873 \\\hline
 NGC3109     & $0.4^{+0.02}_{-0.02}$  & $3.98^{+0.65}_{-0.5}$   & $3.52^{+0.19}_{-0.19}$ & $7.29^{+0.07}_{-0.07}$ & $0.48^{+0.48}_{-0.22}$   & 0.183  \\
 NGC3109     & $0.62^{+0.0}_{-0.0}$   & $5.14^{+0.94}_{-0.71}$  & $2.08^{+0.01}_{-0.0}$  & $7.27^{+0.07}_{-0.07}$ & $0.47^{+0.47}_{-0.22}$   & 0.1824 \\
 NGC3109     & $0.57^{+0.01}_{-0.01}$ & $5.21^{+0.97}_{-0.74}$  & $2.15^{+0.03}_{-0.02}$ & $7.25^{+0.07}_{-0.07}$ & $0.48^{+0.47}_{-0.22}$   & 0.183  \\\hline
 UGC00128    & $0.44^{+0.03}_{-0.03}$ & $7.93^{+0.45}_{-0.44}$  & $3.15^{+0.24}_{-0.22}$ & $7.17^{+0.07}_{-0.07}$ & $0.31^{+0.28}_{-0.14}$   & 0.1696 \\
 UGC00128    & $0.61^{+0.01}_{-0.01}$ & $10.2^{+0.89}_{-0.87}$  & $2.08^{+0.01}_{-0.0}$  & $7.12^{+0.06}_{-0.06}$ & $0.38^{+0.28}_{-0.14}$   & 0.1952 \\
 UGC00128    & $0.55^{+0.02}_{-0.02}$ & $10.12^{+0.99}_{-0.95}$ & $2.21^{+0.05}_{-0.04}$ & $7.11^{+0.07}_{-0.06}$ & $0.35^{+0.28}_{-0.14}$   & 0.1696 \\\hline
 UGC06446    & $0.62^{+0.0}_{-0.01}$  & $2.11^{+0.33}_{-0.33}$  & $2.09^{+0.04}_{-0.01}$ & $8.04^{+0.13}_{-0.12}$ & $0.71^{+0.37}_{-0.18}$   & 0.2966 \\
 UGC06446    & $0.42^{+0.04}_{-0.05}$ & $1.94^{+0.44}_{-0.4}$   & $2.64^{+0.22}_{-0.17}$ & $8.02^{+0.14}_{-0.13}$ & $0.61^{+0.35}_{-0.18}$   & 0.3752 \\
 UGC06446    & $0.32^{+0.04}_{-0.05}$ & $1.81^{+0.43}_{-0.4}$   & $3.14^{+0.31}_{-0.24}$ & $8.0^{+0.15}_{-0.13}$  & $0.52^{+0.35}_{-0.17}$   & 0.2966 \\\hline
 UGC06667    & $0.59^{+0.02}_{-0.03}$ & $3.29^{+0.51}_{-0.44}$  & $2.27^{+0.12}_{-0.09}$ & $7.72^{+0.1}_{-0.1}$   & $0.93^{+1.23}_{-0.62}$   & 0.1258 \\
 UGC06667    & $0.53^{+0.03}_{-0.03}$ & $3.53^{+0.66}_{-0.58}$  & $2.28^{+0.09}_{-0.07}$ & $7.66^{+0.1}_{-0.1}$   & $0.94^{+1.26}_{-0.63}$   & 0.1123 \\
 UGC06667    & $0.43^{+0.03}_{-0.03}$ & $3.48^{+0.67}_{-0.59}$  & $2.6^{+0.13}_{-0.11}$  & $7.63^{+0.1}_{-0.1}$   & $0.96^{+1.26}_{-0.63}$   & 0.1258 \\\hline
 UGC08286    & $0.62^{+0.01}_{-0.01}$ & $1.96^{+0.22}_{-0.21}$  & $2.08^{+0.01}_{-0.0}$  & $8.14^{+0.08}_{-0.08}$ & $1.79^{+0.88}_{-0.43}$   & 0.7475 \\
 UGC08286    & $0.39^{+0.03}_{-0.03}$ & $1.82^{+0.26}_{-0.24}$  & $2.76^{+0.14}_{-0.12}$ & $8.08^{+0.09}_{-0.08}$ & $1.42^{+0.86}_{-0.42}$   & 0.8502 \\
 UGC08286    & $0.3^{+0.03}_{-0.03}$  & $1.75^{+0.26}_{-0.24}$  & $3.27^{+0.18}_{-0.17}$ & $8.05^{+0.09}_{-0.09}$ & $1.3^{+0.87}_{-0.42}$    & 0.7475 \\\hline
 NGC4376     & $0.57^{+0.01}_{-0.01}$ & $1.78^{+0.11}_{-0.11}$  & $2.17^{+0.02}_{-0.01}$ & $8.32^{+0.03}_{-0.03}$ & $1.87^{+0.3}_{-0.15}$    & 1.394  \\
 NGC4376     & $0.32^{+0.01}_{-0.01}$ & $1.74^{+0.12}_{-0.12}$  & $3.14^{+0.05}_{-0.05}$ & $8.22^{+0.03}_{-0.03}$ & $1.72^{+0.3}_{-0.15}$    & 1.5238 \\
 NGC4376     & $0.24^{+0.01}_{-0.01}$ & $1.75^{+0.12}_{-0.12}$  & $3.71^{+0.06}_{-0.05}$ & $8.17^{+0.03}_{-0.03}$ & $1.59^{+0.3}_{-0.15}$    & 1.394  \\\hline
 NGC4396     & $0.6^{+0.0}_{-0.01}$   & $2.19^{+0.12}_{-0.12}$  & $2.1^{+0.01}_{-0.01}$  & $8.18^{+0.03}_{-0.03}$ & $1.86^{+0.09}_{-0.05}$   & 1.3254 \\
 NGC4396     & $0.36^{+0.01}_{-0.01}$ & $2.07^{+0.12}_{-0.12}$  & $2.92^{+0.05}_{-0.05}$ & $8.1^{+0.03}_{-0.03}$  & $1.58^{+0.09}_{-0.05}$   & 1.514  \\
 NGC4396     & $0.27^{+0.01}_{-0.01}$ & $2.03^{+0.12}_{-0.13}$  & $3.47^{+0.06}_{-0.06}$ & $8.06^{+0.03}_{-0.03}$ & $1.39^{+0.09}_{-0.05}$   & 1.3254 \\\hline
 NGC7320     & $0.45^{+0.01}_{-0.01}$ & $1.03^{+0.06}_{-0.06}$  & $2.54^{+0.04}_{-0.04}$ & $8.71^{+0.03}_{-0.03}$ & $1.75^{+0.16}_{-0.08}$   & 1.2857 \\
 NGC7320     & $0.21^{+0.01}_{-0.01}$ & $0.95^{+0.06}_{-0.06}$  & $4.01^{+0.07}_{-0.07}$ & $8.62^{+0.04}_{-0.04}$ & $1.45^{+0.16}_{-0.08}$   & 1.3416 \\
 NGC7320     & $0.16^{+0.01}_{-0.01}$ & $0.94^{+0.07}_{-0.07}$  & $4.6^{+0.07}_{-0.07}$  & $8.59^{+0.04}_{-0.04}$ & $1.39^{+0.16}_{-0.08}$   & 1.2857 \\\hline
 UGC04169    & $0.62^{+0.0}_{-0.0}$   & $2.52^{+0.18}_{-0.17}$  & $2.08^{+0.0}_{-0.0}$   & $8.03^{+0.05}_{-0.04}$ & $1.8^{+0.18}_{-0.09}$    & 1.6352 \\
 UGC04169    & $0.42^{+0.01}_{-0.01}$ & $2.48^{+0.2}_{-0.19}$   & $2.66^{+0.06}_{-0.06}$ & $7.95^{+0.04}_{-0.04}$ & $1.75^{+0.18}_{-0.09}$   & 1.6288 \\
 UGC04169    & $0.32^{+0.01}_{-0.01}$ & $2.47^{+0.2}_{-0.2}$    & $3.12^{+0.08}_{-0.08}$ & $7.91^{+0.04}_{-0.04}$ & $1.76^{+0.18}_{-0.09}$   & 1.6352 \\\hline
    \end{tabular}
    \caption{From the left to right columns: galaxy name, the core radius normalized to the initial scale radius ($r_c/r_{s,0}$), the core radius ($r_c$), the core density normalized to the initial scale density ($\rho_c/\rho_{s,0}$), the core density $\rho_c$, $\chi^2$ per degrees of freedom and its minimal value. For each galaxy, the first, second, and third rows are for $\sigmam=3~\cmg$, $20~\cmg$, and $40~\cmg$, respectively.}
    \label{tab:numerics2}
\end{table}{}

\newpage

\section{Detailed fits to the rotation curves in the sample}

\label{sec:app-rotation-curves}

In Fig.~\ref{fig:ugc08286}, we show detailed fits to the rotation curves for all galaxies in our sample based on the parametric model with $\sigmam=3~\cmg$ (left), $20~\cmg$ (middle), and $40~\cmg$ (right). The mass-to-light ratio is fixed to $\Upsilon_{\star,\rm{disk}} = 0.5~ M_{\odot}/L_{\odot}$. As discussed in Sec.~\ref{sec:galaxy-sample}, the galaxies in our sample are dark matter-dominated in their central regions, and we can see this explicitly by examining each contribution of the total rotation curve.

\begin{figure}[h]
\scalebox{0.7}{\includegraphics{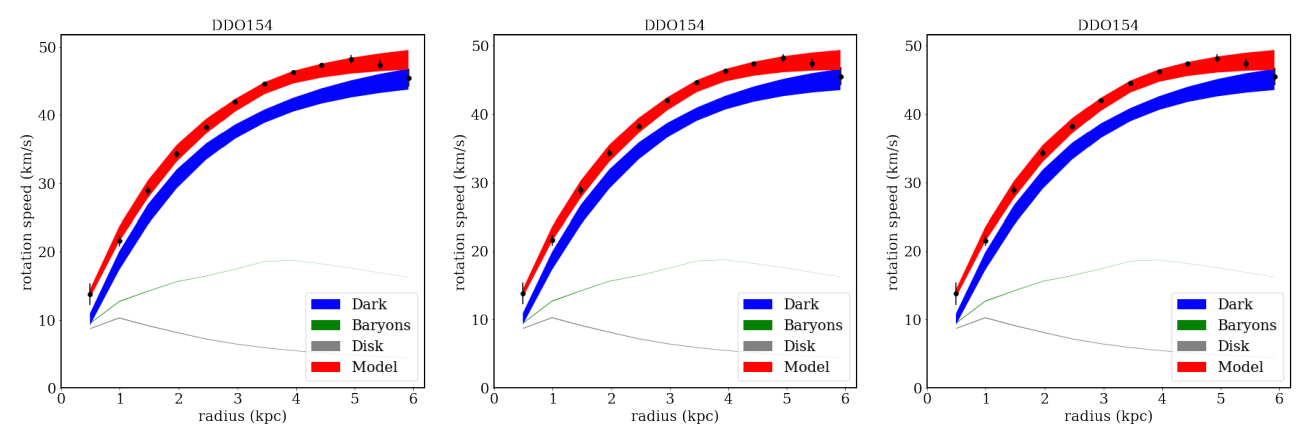}}
\label{fig:ddo154}

\scalebox{0.7}{\includegraphics{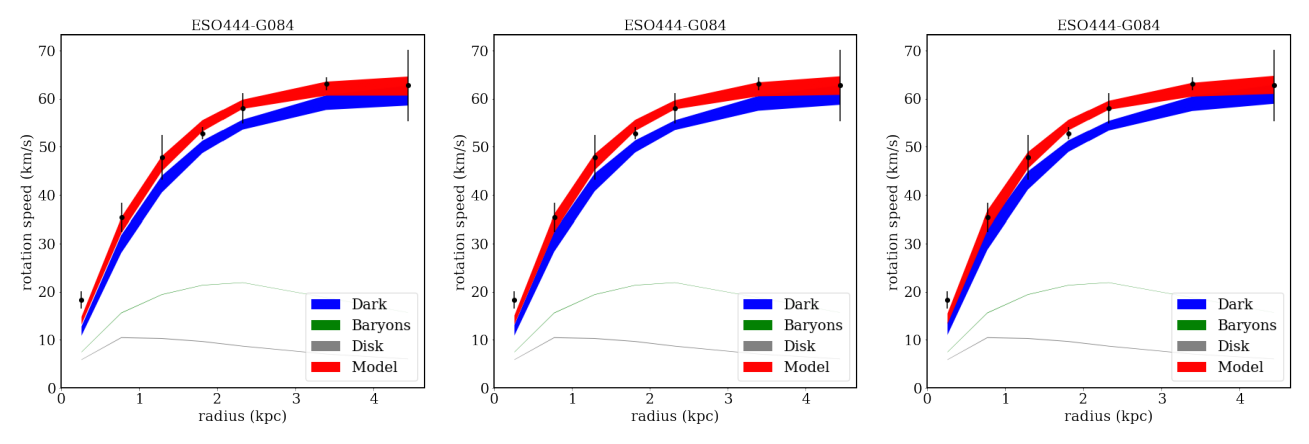}}
\label{fig:eso444}

\scalebox{0.7}{\includegraphics{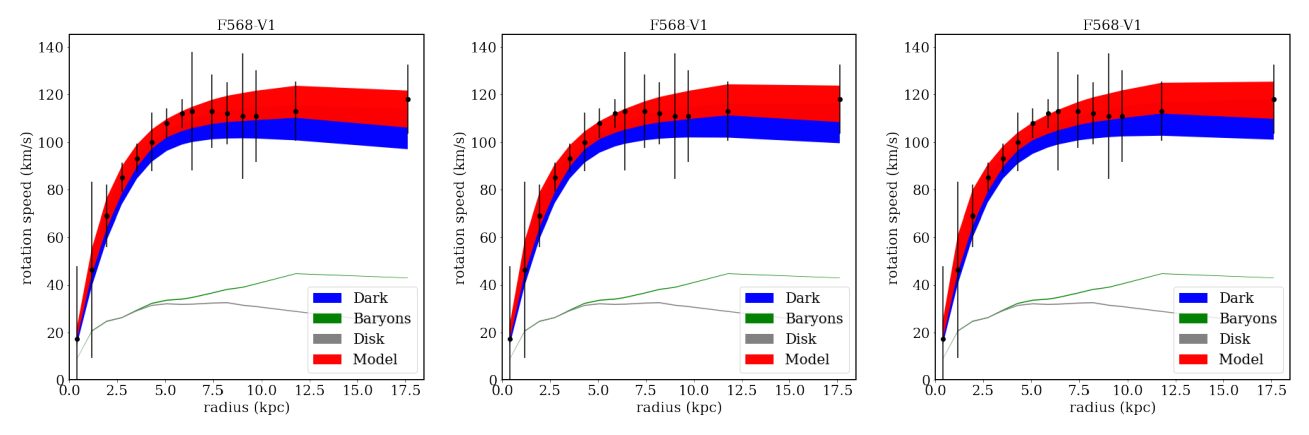}}
\label{fig:f568-v1}

\scalebox{0.7}{\includegraphics{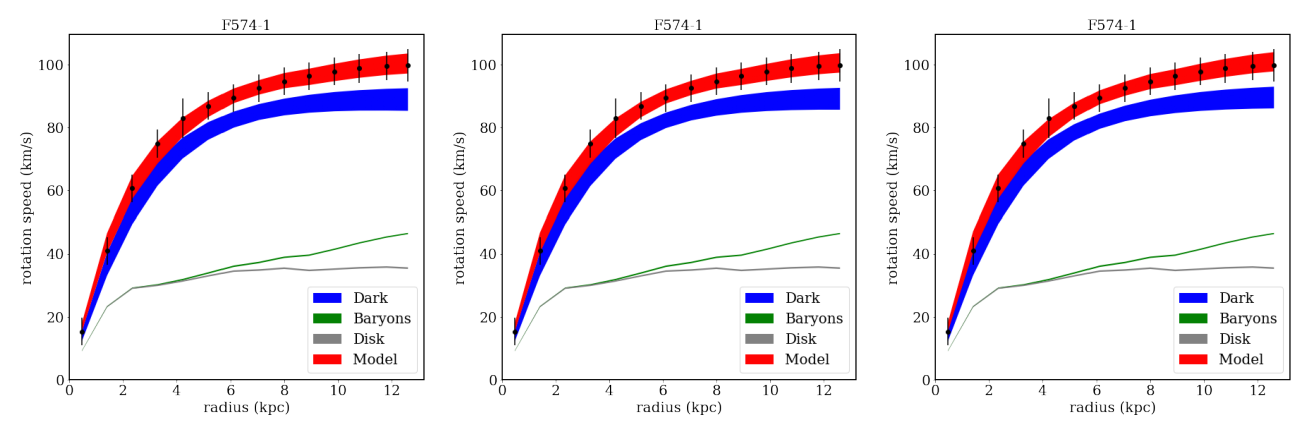}}
\label{fig:f574-1}
\end{figure}

\newpage

\begin{figure}[h]
\scalebox{0.7}{\includegraphics{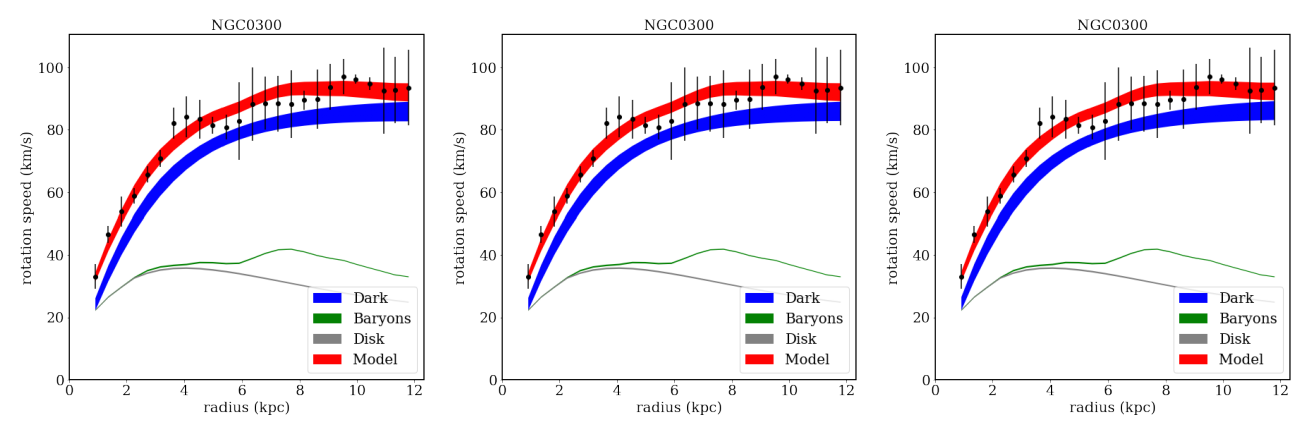}}
\label{fig:ngc300}

\scalebox{0.7}{\includegraphics{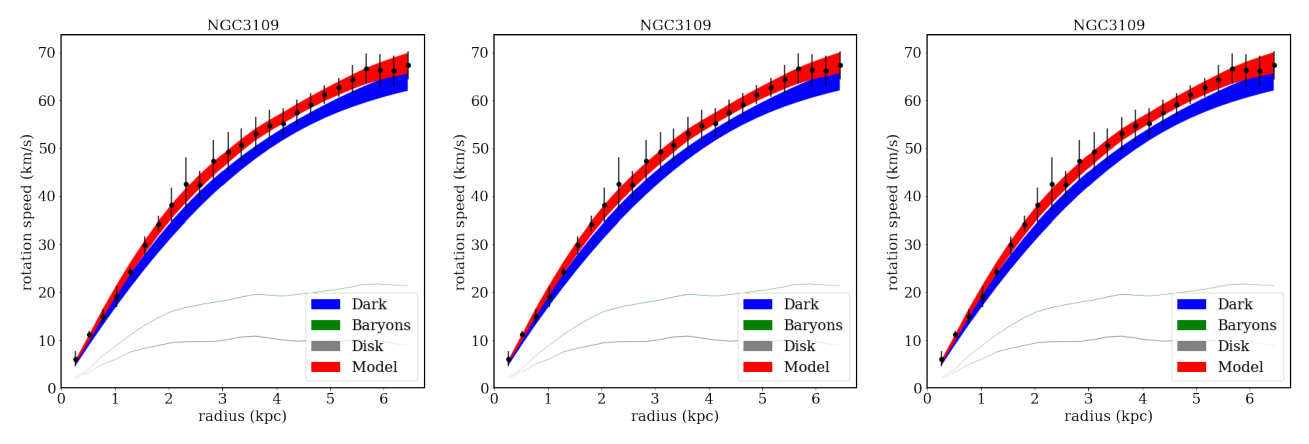}}
\label{fig:ngc3109}

\scalebox{0.7}{\includegraphics{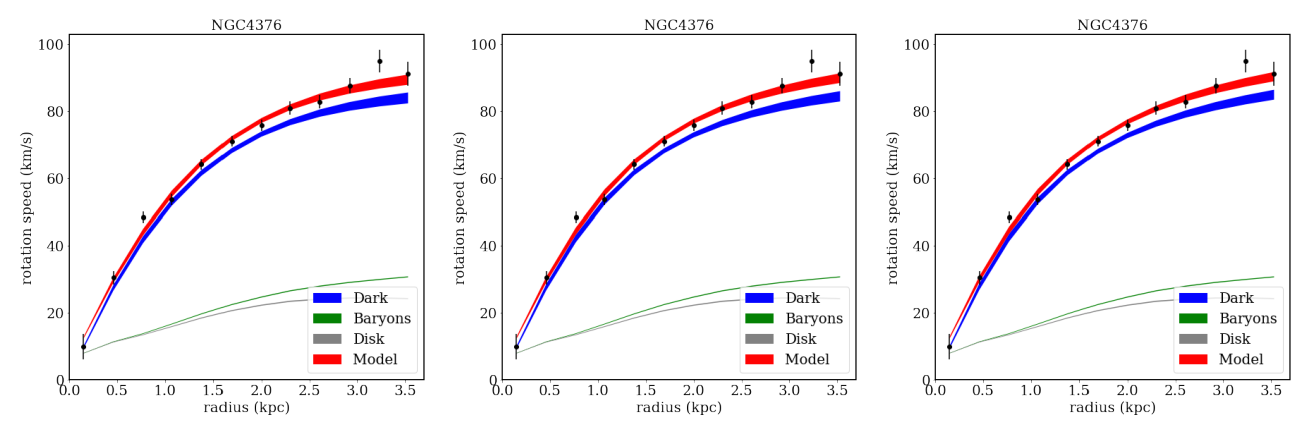}}
\label{fig:ngc4376}

\scalebox{0.7}{\includegraphics{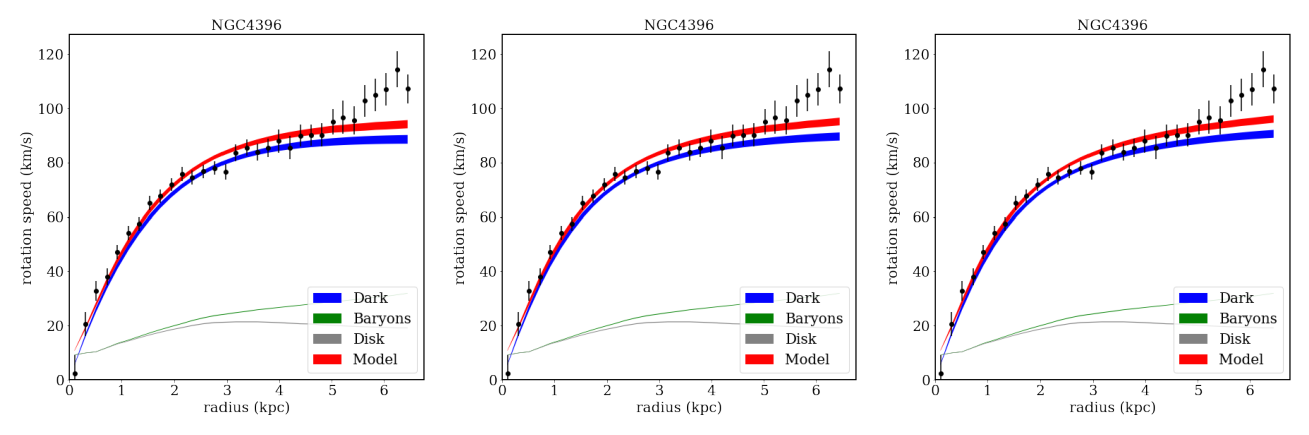}}
\label{fig:ngc4396}
\end{figure}

\newpage

\begin{figure}[h]
\scalebox{0.7}{\includegraphics{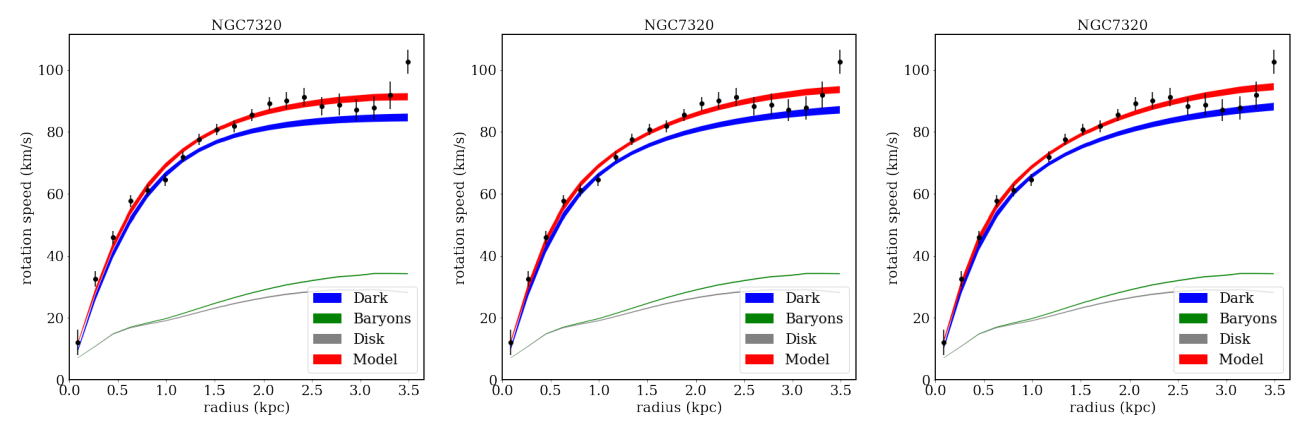}}
\label{fig:ngc7320}

\scalebox{0.7}{\includegraphics{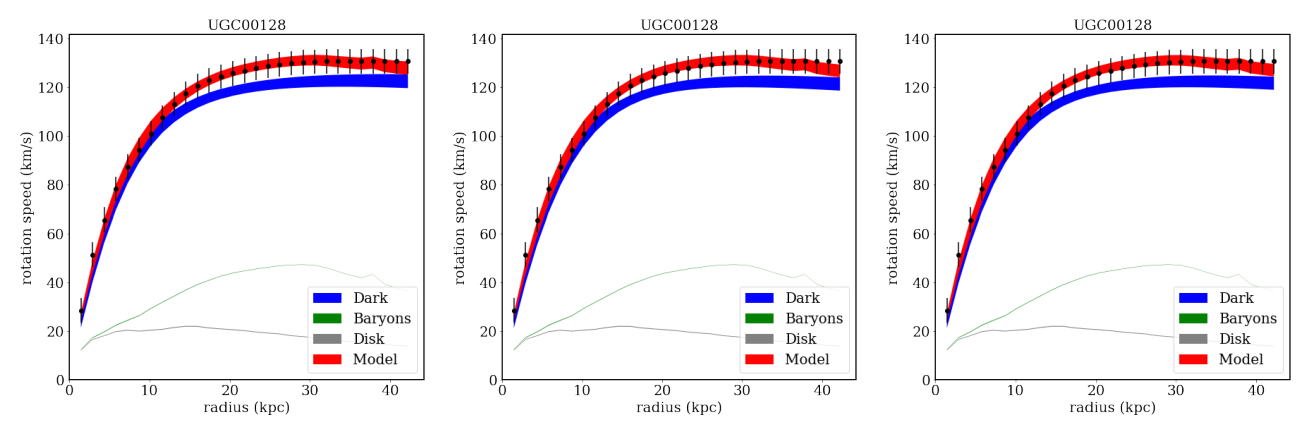}}
\label{fig:ugc00128}

\scalebox{0.7}{\includegraphics{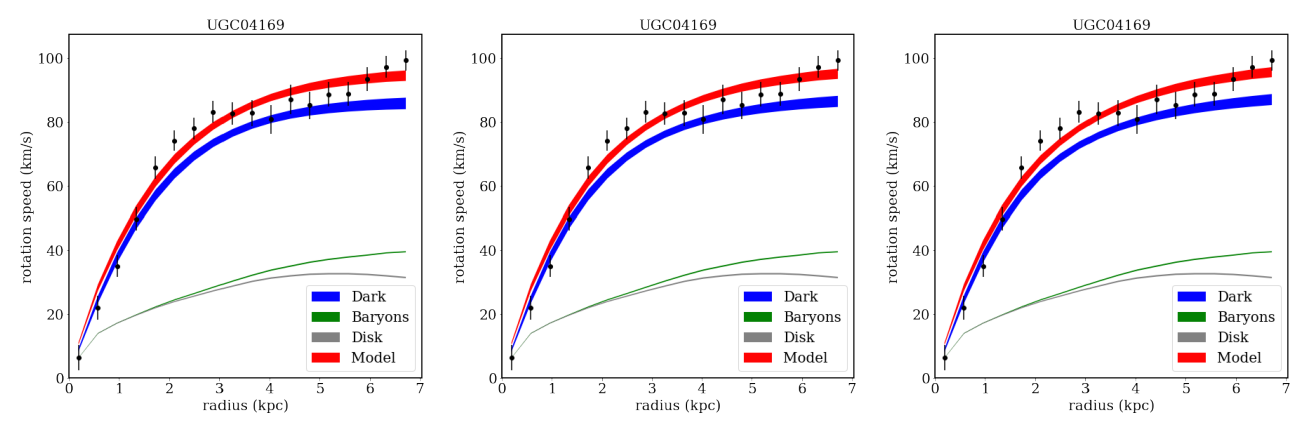}}
\label{fig:ugc04169}

\scalebox{0.7}{\includegraphics{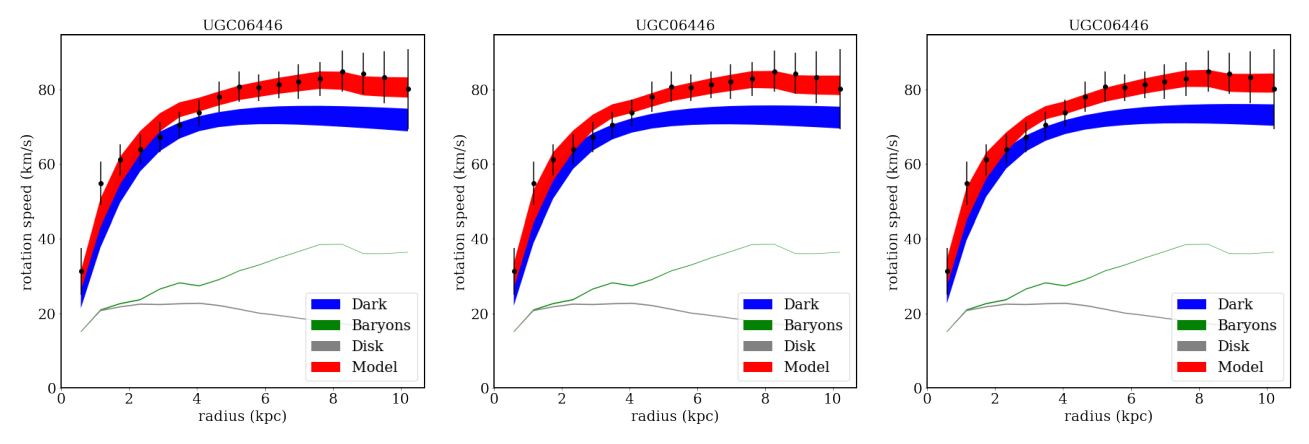}}
\label{fig:ugc06446}
\end{figure}

\newpage

\begin{figure}[h]
\scalebox{0.7}{\includegraphics{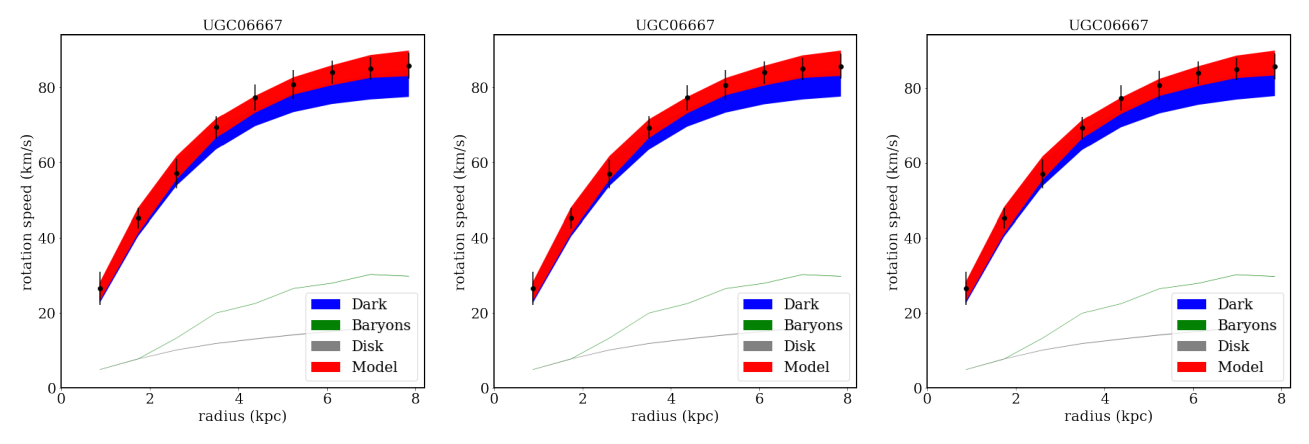}}
\label{fig:ugc06667}

\scalebox{0.7}{\includegraphics{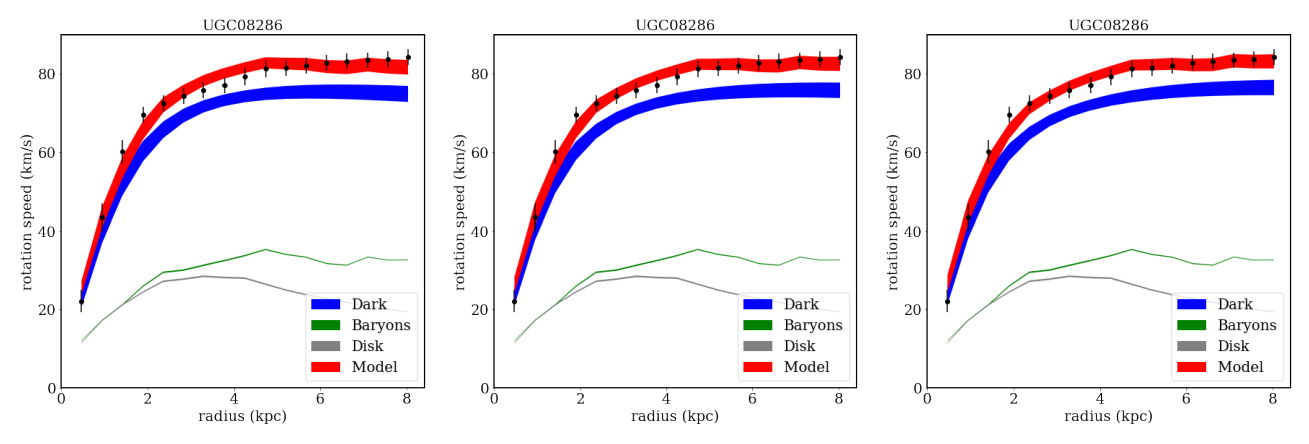}}
\caption{The SIDM fits to the rotation curves of the spiral galaxies in our sample based on the parametric model with $\sigmam = 3~\cmg$ (left), $20~\cmg$ (middle), and $\sigmam = 40~\cmg$ (right), including the  total circular velocity from the model prediction (solid-red), the halo concentration (solid-blue), the total baryon concentration (solid-green), as well as one from the stellar disk (solid-gray).}
\label{fig:ugc08286}
\end{figure}

\newpage

\end{document}